\documentclass[12pt,preprint]{aastex}

\slugcomment{Not to appear in Nonlearned J., 45.}

\shorttitle{AGN-Host Connection in Partially Obscured AGNs. I}
\shortauthors{Wang et al.}

\begin{document}


\title{Understanding AGN-Host Connection in Partially Obscured Active Galactic Nuclei.\\
Part I: The Nature of AGN+HII Composites \\ }


\author{J. Wang and J. Y. Wei}
\affil{National Astronomical Observatories, Chinese Academy of Science, 20A Datun Road, 
Chaoyang District, Beijing 100012, China}

\email{wj@bao.ac.cn}




\begin{abstract}
The goal of our serial papers is to examine the evolutionary connection between AGN and
star formation in its host galaxy in the partially obscured AGNs (i.e., Seyfert 1.8 and 1.9 galaxies). 
Taking advantage of these galaxies, the properties of both components can be studied
together by direct measurements. In this paper, we focus on the broad-line composite galaxies
(composite AGNs) which are located between the theoretical and empirical separation lines in
the [\ion{N}{2}]/H$\alpha$ vs. [\ion{O}{3}]/H$\beta$ diagram. These galaxies are searched for 
from the composite galaxies provided by the SDSS DR4 MPA/JHU catalogs. After re-analyze the spectra, 
we perform a fine classification for the 85 composite AGNs in terms of the BPT diagrams.
All the objects located below the three theoretical separation lines are associated with
a young stellar population ($<$1Gyrs), while either a young or old stellar population is identified
in the individual multiply-classified object. 
The multiply-classified objects with a very old stellar population 
are located in the LINER region in the [\ion{O}{1}]/H$\alpha$ vs. [\ion{O}{3}]/H$\beta$ diagram. 
We then consider the connection between AGN and star formation to derive the key results.  
The Eddington ratio inferred from the broad H$\alpha$ emission,
the age of the stellar population of AGN's host as assessed by $D_n(4000)$, and the line ratio
[\ion{O}{1}]/H$\alpha$ are found to be related with each other. These relations strongly support 
the evolutionary scenario in which AGNs evolve from high $L/L_{\mathrm{Edd}}$ state with soft spectrum 
to low $L/L_{\mathrm{Edd}}$ state with hard spectrum as young stellar population ages and fades. 
The significant correlation between the line ratio
[\ion{O}{1}]/H$\alpha$ and $D_n(4000)$ leads us to suggest that the
line ratio could be used to trace the age of stellar population in type I AGNs.

\end{abstract}





\section{Introduction}
\subsection{Composite galaxies}
The Baldwin-Phillips-Terlevich (BPT) diagrams (Baldwin et al. 1981) are commonly used as a powerful
tool to classify narrow emission-line galaxies according to their dominant energy sources.
The diagrams are then refined by Veilleux \& Osterbrock (1987). In the diagnostic diagram
[\ion{O}{3}]$\lambda5007$/H$\beta$ vs. [\ion{N}{2}]$\lambda6583$/H$\alpha$,
the star-forming sequence shows a curved concentration extending from up-left
to bottom-right, while AGNs (including LINERs, Low Ionization Nuclear Emission-line Regions)
concentrate on the right part of the diagram. A theoretic demarcation curve determining the upper limit
of the theoretical stellar photoionization model is proposed by Kewley et al. (2001),
and an empirical curve dividing the pure star-forming galaxies recently by Kauffmann et al. (2003a). The
region between the two curves is occupied by the so-called composite objects (hereafter composites
for short) whose spectra are believed
to contain significant contributions from both star formation and AGN (Seyfert or LINER).

As a fundamental, tight relationship, the well-established relation between the
mass of supermassive black hole (SMBH) and velocity dispersion of 
the bugle in which the SMBH resides is served as a clue of co-evolution of 
AGN and star formation in the bugle 
(e.g., Magorrian et al. 1998; Gebhardt et al. 2000; Granato et al. 2001;  
Tremaine et al. 2002; Ferrarese \& Merritt 2001; Greene \& Ho 2006). At first, 
the issue of co-evolution could be 
studied in phenomenology. In fact, there are many observations of both imaging and 
spectroscopy that show the evidence of  
co-existence of AGN and circumnuclear star formation (e.g., Canalizo \& Stockton 2001; 
Cid Fernandes et al. 2001, 2005; Hines et al. 1999; Schinerer et al. 1998; 
Heckman et al. 1997; Gonzalez Delgado 2002; Zuther et al. 2007; Sarzi et al. 2007; Asari et al. 2007;
Davies et al. 2007). 

We should next answer the 
questions: What is the physics that links star formation and AGN phenomena? How AGN co-evolves 
with star formation in details? (e.g., Ho 2005a). So far, 
there are two distinct scenarios that connect AGN and 
star formation phenomena: 1) starburst occurs first and triggers AGN's activity latter (e.g., 
Weedman 1983); 
2) in contrast, AGN's activity occurs before the beginning of starburst 
(Goncalves et al. 1999 and references therein). 
If the hypothesis of co-evolution of AGN and the bulge of its host 
is correct, then whatever AGN occurs before/after
starburst, the composites might represent an important state in galaxy evolution,
and are valuable for examining the issue of co-evolution in details.

\subsection{A new approach by Seyfert 1.8 and 1.9 galaxies}

In order to study the co-evolution of AGN and star formation, properties of both     
AGN and its host galaxy are required to be examined carefully and simultaneously. 
However, discrimination between type I AGN and its host galaxy is a difficult 
task, because both the AGN's continuum and broad lines mask the stellar features in optical 
wavelength. The problem of AGN-host separation could be partially solved by several 
techniques listed as follows: 
\begin{itemize}
\item Off-nuclear spectroscopy: By detecting the Balmer absorption lines and emission 
lines of \ion{H}{2} region in the surrounding of QSOs, 
Canalizo \& Stockton (2001) examined properties of circumnuclear stellar 
populations in the hosts of nine local QSOs. The estimated ages range from a few Myrs
old to post-starburst. In a similar way, Nolan et al. (2001) found that the off-nuclear 
($\sim$5\symbol{125}) stellar populations of optically selected QSOs are generally
dominated by old ones ($\sim$8-14Gyrs). Relatively old stellar populations ($\sim$0.1-2Gyrs) 
are also frequently detected in the off-nuclear regions in a few powerful radio-loud AGNs 
(Tadhunter et al. 2005).
\item AGN+post starburst: In this case the star light is easily discriminated in the 1-D spectroscopy 
because of the 
prominent Balmer absorptions of A-type stars. Brotherton et al. (1999) identified a 400Myrs old 
instantaneous starburst in quasar \object{UN\,J1025-0040}. A $\sim100$Myrs old post-starburst is
identified in NLS1-like type I AGN \object{SDSS\,J085338.27+033246.1} (Wang \& Wei 2006).
Wang et al. (2004) identified an evident Balmer jump in type I AGN 
\object{SDSS\,J022119.84+005628.4}. 
\item Spectroscopy of obscured AGNs: By analyzing the spectra of narrow-line AGNs from Sloan Digital 
Sky Survey (SDSS; York et al. 2000),  Heckman et al. (2004 and references therein) 
and Kewley et al. (2006) 
found that most of the accretion-driven growth of mass of SMBH occurs in the galaxy associated with a relatively 
young stellar population (see also in Heckman \& Kauffmann 2006).   
\item Infrared color: By analyzing a sample of IRAS-selected type 1.5 AGNs,
Wang et al. (2006) found that the essential Eigenvector 1 (E1) space 
(Boroson \& Green 1992, hereafter BG92) is related with the age of stellar population of 
the host as assessed by infrared color $\alpha(60,25)$.
\item Deconvolution of 2-D spectra: By performing the MCS deconvolution method on 2-D spectra,
Letawe et al. (2007, and references therein) recently separated the individual spectra of the 
quasar and of the underlying host galaxy in a sample 
of 20 low-redshift luminous quasars. Quite young stellar populations ($\sim1-2$ Gyrs) are subsequently  
identified in a sample of 18 quasars by a semi-analytic decomposition (Jahnke et al. 2007).   
\end{itemize}

Although the understanding of the co-evolution of AGN and star formation 
has been remarkably developed over the past few years by using these approaches, 
their intrinsic dis-advantages prevent the systematical and 
extensive investigation on the issue of co-evolution. 
We refer the readers to Letawe et al. (2007) for a comment of the drawbacks of these
approaches. In the present study, taking advantage of Seyfert 1.8 and 1.9 galaxies (i.e., partially obscured AGNs),
a new approach is adopted by us to directly obtain
properties of both central AGN and circumnuclear stellar population systematically and 
simultaneously. Seyfert 1.8 galaxy is defined as one whose spectrum shows strong narrow components, 
and very weak but still visible broad components of H$\alpha$ and H$\beta$. 
No broad components of Balmer lines except a weak broad component at H$\alpha$ can be identified in
Seyfert 1.9 galaxy (Osterbrock 1989).
In these two types of AGN,
due to the orientation effect (e.g., Antonucci 1993), their optical continuum emitted from the central accretion 
disks is blocked by a torus, and the star light from the bugle is much easier to be detected. 
In addition, the broad emission lines shown in the spectrum, rather than some tracers, allow us to derive 
AGN's properties directly.

As the first step, by deriving the properties of both AGN and stellar population
from the optical spectra taken by SDSS, we focus on the nature of 
the composites with broad Balmer emission lines (hereafter composite
AGNs for abbreviation), and then study their implications on the issue of co-evolution
in this paper.
The issues related with other types of galaxies (e.g., Seyfert and LINER) 
will be presented in subsequent paper(s).
At present, SDSS provides a large and uniform spectroscopic 
sample of galaxies to make such study available. The paper is organized as follows. 
The sample selection and data reduction are
described in Section 2. The physical properties of both AGN and stellar population are 
derived in Section 3. 
Basing upon the BPT diagrams, a further classification of the composite AGNs is shown in Section 4. 
We investigate the AGN-host connection in the composite AGNs in Section 5.
The $\Lambda$ cold dark matter ($\Lambda$CDM) cosmology 
with parameter $h=0.7$, $\Omega_{\rm M}=0.3$, and $\Omega_{\Lambda}=0.7$ (Spergel et al. 2003)
is adopted throughout the paper.

\section{Sample Selection and Data Reduction}

\subsection{The Sloan Digital Sky Survey}
The SDSS is designed to eventually provide an imaging and spectroscopic survey of 
one-quarter of the entire sky (e.g., York et al. 2000). 
The survey is carried out by a dedicated 2.5-meter wide-field (3\symbol{23})
telescope at Apache Point Observatory. The telescope is equipped  
with two fiber-fed spectrographes and a mosaic CCD camera. 
Each spectrum is taken with a 3\symbol{125} diameter fiber aperture, and covers a wavelength range 
from 3800\AA\ to 9200\AA\ at the observe-frame. The spectral resolution varies from 1850 to 2200, 
corresponding to $\sigma_{\rm{int}}\sim65 \rm{km\ s^{-1}}$. The spectrophotometric calibrations
with accuracy of about 20\%
are performed by observing subdwarf F stars in each 3\symbol{23} field of view.
Tow pipelines, \sl spectro2d \rm and \sl specBS \rm are developed to reduce 
the raw spectra (Glazebrook et al. 1998; Bromley et al. 1998).

\subsection{Selection of Seyfert 1.8 and 1.9 galaxies}

We select a sample of the composite AGNs from the SDSS-Data Release 
4 (Adelman-McCarthy et al. 2006) MPA/JHU catalogs (Heckman et al. 2004; Kauffmann et al. 2003a,b,c; 
Tremonti et al., 2004)\footnote{These catalogs can be 
downloaded from http://www.mpa-garching.mpg.de/SDSS/.}. 
The catalogs provide spectral and physical properties for 567,486 individual 
narrow emission-line galaxies, including Seyferts, LINERs and star-forming galaxies.  

First of all, in order to ensure that our results are not distorted by the spectra 
with poor quality, we only consider 
the galaxies whose spectra have median signal-to-noise ratio per pixel of the whole spectrum 
$\rm{S/N}\geq20$. Secondly, the star-forming galaxies which are under the demarcation line given in
Kauffmann et al. (2003a) are excluded by simply adopting the emission 
line intensities provided by the MPA/JHU catalogs. The extinction is not corrected here
since both the line ratios [\ion{O}{3}]/H$\beta$ and [\ion{N}{2}]/H$\alpha$ are not sensitive to
dust extinction. There are total 27,341 galaxies that fulfill these criterion.
An object is then classified as a composite if  
\begin{equation}
\rm{\log([OIII]/H\beta)<0.61/(\log([NII]/H\alpha)-0.05)+1.3}
\end{equation}
and 
\begin{equation}
\rm{\log([OIII]/H\beta)>0.61/(\log([NII]/H\alpha)-0.47)+1.19}
\end{equation}
Altogether, there are 12,343 objects that could be classified as the composites with median S/N 
ratio lager than 20.

An automatic routine is then developed by us to reduce the raw data and to select 
objects with broad emission lines attributed to central AGN. At the beginning, 
the Galactic extinction is corrected for each spectrum by assuming an $R_V=3.1$ 
extinction law (Cardelli et al. 1989). The color excess $E(B-V)$ for each spectrum 
is taken from the Schlegel, Finkbeiner and Davis Galactic reddening maps 
(Schlegel et al. 1998). Each of the extinction-corrected spectra is then 
transformed to the rest frame, along with the $k$-correction, 
according to the corresponding redshift provided by the SDSS pipelines.

The contribution of star light is then separated from each
spectrum by the principle component analysis (PCA) method 
(e.g., Yip et al. 2004; Li et al, 2005; Hao et al. 2005). A library of stellar absorption 
spectra is built by applying the PCA technique to the standard 
single stellar population (SSP) models developed by Bruszual \& Charlot 
(2003, hereafter BC03). The models have a resolution of 3\AA.    
The optical spectral evolution is provided for six different metallicities.
At each metallicity, there are total 221 spectra at the unequally
spaced time steps from the age of 0.1Myrs to 20 Gyrs.  
The first seven eigenspectra are used to 
model the starlight component simply through their linear combination. 
The component of AGN's continuum is ignored in the separation in most of the cases,
since the contribution of the observed continuum of AGN is negligible in their spectra (e.g., Kauffmann et al. 2003a),
except in 7 objects. In summary, in order 
to appropriately separate the starlight, our template contains the seven 
eigenspectra and a Galactic extinction curve (Cardelli et al. 1989) for 
most of the cases. Because only the instrumental resolution is taken into 
account in the models of BC03, the template is convolved with a Gaussian 
function to match the velocity dispersion in each object. The velocity dispersion 
is determined by the cross-correlation method before the fitting and subtraction.  
For the cases that show emission of both broad H$\beta$
line and \ion{Fe}{2} complex of AGN, both an additional 
power-law continuum with form $F_\nu\propto\nu^{-0.5}$ (Richards et al. 2006; Elvis et al. 1994)
and a template of the \ion{Fe}{2} complex are required to model the spectra. We 
simply adopt the empirical template of the \ion{Fe}{2} complex given in BG92. The line
width of the template is forced to be identical to that of H$\beta$ broad component (e.g., BG92). 

A $\chi^2$ minimization is used to model the starlight component in each spectrum. 
The fitting is performed over the rest frame wavelength 
range from 3700\AA\ to 7000\AA, except for the regions around the strong emission lines:
e.g., Balmer lines, [\ion{O}{3}]$\lambda\lambda4959,5007$, [\ion{N}{2}]$\lambda\lambda6548,6583$,
[\ion{S}{2}]$\lambda\lambda6716,6731$, [\ion{O}{2}]$\lambda3727$.
The possibly broad components of Balmer lines are also considered to determine
the emission-line free regions. As an illustration, the modeling of the starlight 
component is shown in the left column of Figure 1 for three typical cases. 
In each panel, the upper and lower curves shows the 
starlight-subtracted (emission-line) spectrum and modeled starlight spectrum, respectively.

The emission-line spectra are subsequently used to search for the composite AGNs, 
according to the existence of a high velocity wing of 
broad H$\alpha$ component at its blue side. The red wing is not 
used in the sample selection because of the superposition of the strong [\ion{N}{2}]$\lambda$6583 
emission line.
The existence of a broad wing is accepted 
if the flux ratio $F_{\rm{w}}/\sigma_{\rm c}\geq3$, where $F_{\rm{w}}$ is the
flux of line wing averaged over the wavelength range from 6500\AA\ to 6530\AA\  
at the rest frame, $\sigma_{\rm c}$ the standard deviation of flux within the emission-line
free region between $\lambda$5980 and $\lambda$6020. Iterative tests on 
a typical small sample taken from SDSS are performed to determine 
the wavelength ranges quoted above. The automatically selected composite AGNs are then inspected 
by eyes. Finally, our sample totally contains 85 composite AGNs with broad H$\alpha$ emission 
line. 
 
\subsection{Emission line measurements}

The emission lines are measured once the starlight component is subtracted. 
We model and measure emission lines of the 85 composite AGNs by the SPECFIT task (Kriss 1994) in the IRAF 
package\footnote{IRAF is distributed by the National Optical Astronomical Observatories,
which is operated by the Association of Universities for Research in Astronomy, Inc.,
under cooperative agreement with the National Science Foundation.}. 
In each spectrum, each emission line is fitted by a set of Gaussian components. The intensity 
ratios of the [\ion{O}{3}] and [\ion{N}{2}] doublets are fixed to their theoretical values (e.g., 
Dimitrijevic et al. 2007). 
In Figure 1, the emission-line modeling in H$\beta$ region is schemed in the middle column, 
and the modeling in H$\alpha$ region in the right column. 
Direct integration is used to measure the flux 
of [\ion{O}{1}]$\lambda6300$ emission line in each spectrum through the SPLOT task. 
The uncertainties 
of the emission-line measurements are dominated by the subtraction of the starlight component, especially 
for the faint H$\beta$ emission. Similar as Asari et al. (2007), a wide ($\sim200$\AA) absorption trough 
in the continuum around H$\beta$ is frequently identified in our starlight-subtracted spectra, especially 
in the galaxies associated with an old stellar population, which leads to a 
2\%-7\% underestimate in H$\beta$ flux if the continuum is fixed to be zero. 
This small mismatch is probably caused by the calibration in the STELIB library (Asari et al. 2007). 
We ignore this small underestimate in this paper 
because relatively young stellar populations are frequently identified in our sample 
(see Figure 5 for details).

The results of the emission-line measurements are
listed in Table 1. Column (1) lists the identification of each object listed in our sample,
and Column (2) the corresponding redshift. For each object, the line ratios of [\ion{N}{2}]/H$\alpha$, 
[\ion{S}{2}]$\lambda\lambda6726,6731$/H$\alpha$, [\ion{O}{1}]/H$\alpha$ and [\ion{O}{3}]/H$\beta$
are listed from Column (3) to Column (6). Column (9) and (12) shows the intrinsic luminosity 
of H$\alpha$ broad component ($L_{\mathrm{H\alpha}}$) and that of [\ion{O}{3}]$\lambda$5007 
($L_{\mathrm{[OIII]}}$), respectively.
Both luminosities are corrected for local extinction. 
The extinction is inferred from the narrow-line ratio H$\alpha$/H$\beta$ for each object, assuming the
Balmer decrement for standard case B recombination and Galactic extinction curve with $R_V=3.1$.

\section{Deriving Physical Properties}

In this section we derive physical properties of both AGN and stellar population
basing upon the reduction and measurements described above. 

\subsection{Deriving properties of AGN}

There are compelling evidence suggesting that both Eddington ratio ($L/L_{\rm E}$) and virial mass of SMBH
($M_{\rm{BH}}$) are basic parameters determining the properties of individual AGN (e.g., Boroson 2002; BG92; 
Sulentic et al. 2000; Grupe 2004). In this study, we attempt to directly obtain the physical 
properties of individual central AGN according to its broad emission lines.
Basing upon the calibration of the measurements of the reverberation mapping methods,
the empirical relationships that are commonly used to estimate $L/L_{\rm E}$ and $M_{\rm{BH}}$ in 
typical type I AGNs have been well established in recent years 
(e.g., Kaspi et al., 2000, 2005; Peterson et al. 2004; 
Vestergaard et al. 2002; Wu et al. 2004). Both velocity of broad H$\beta$ emission line
and luminosity at 5100\AA\ (or luminosity of broad H$\beta$ emission) emitted from the central AGN
are required in these calibrations,
which are however not feasible in our sample both because of the 
dilution of AGN's luminosity 
by the host's starlight and because of the non-detectable broad H$\beta$ emission in a majority of cases.
Greene \& Ho (2005) recently extended 
the estimations of both bolometric luminosity and $M_{\rm{BH}}$ to broad H$\alpha$ line
\begin{equation}
L_{5100}=2.4\times10^{43}\bigg(\frac{L_{\mathrm{H}\alpha}}{10^{42}\ \mathrm{erg\ s^{-1}}}\bigg)^{0.86}\ \mathrm{erg\ s^{-1}}
\end{equation}
\begin{equation}
M_{\mathrm{BH}}=2\times10^{6}\bigg(\frac{L_{\mathrm{H\alpha}}}{10^{42}\ \mathrm{ergs\ s^{-1}}}\bigg)^{0.55}
\bigg(\frac{\mathrm{FWHM(H\alpha)}}{1000\ \mathrm{km\ s^{-1}}}\bigg)^{2.06}M_{\odot}
\end{equation}
The H$\alpha$ emission line provides much more robust estimations of both $L/L_{\rm E}$ and $M_{\rm{BH}}$ 
in the cases where the contamination caused by host and/or non-thermal jet radiation is present.
The relationships allow us to estimate both $L/L_{\rm E}$ and $M_{\rm{BH}}$ in 
the composite AGNs. For each object,
the estimated values of $M_{\rm{BH}}$ and $L/L_{\rm E}$ is listed in Column (10) and 
(11) in Table 1, respectively.

According to the unified model of AGN (Antonucci 1993), the existence of the broad emission lines 
in AGN's spectrum depends
on the orientation of the axis of tours with respect to the line-of-sight of observer. In order to assess
how much the orientation effect affects $L_{\mathrm{H\alpha}}$ (and hence $L/L_{\mathrm{Edd}}$
and $M_{\mathrm{BH}}$)  
that is estimated from the broad H$\alpha$ line, the $L_{[\mathrm{OIII}]}$ vs. 
$L_{\mathrm{H\alpha}}$ diagram is shown in the \sl Panel D \rm in Figure 7.
A strong correlation is found between $L_{[\mathrm{OIII}]}$
and $L_{\mathrm{H\alpha}}$. [\ion{O}{3}] emission is isotropic in AGN because the line is
emitted only from the optically thin region that is at large enough distance from the 
central black hole. The
isotropy of [\ion{O}{3}] emission is also supported by the correlation between [\ion{O}{3}] and
orientation-independent [\ion{O}{2}] emission (Kuraszkiewicz et al. 2000),
although the isotropy
has been questioned in some radio-loud AGNs (Baker 1997; di Serego Alighirei et al. 1997).
The tight correlation shown in the \sl Panel D \rm in Figure 7 therefore indicates that 
the variation of $L_{\mathrm{H\alpha}}$
in our sample is not entirely due to the orientation effect, 
and that $L_{\mathrm{H\alpha}}$ is a good measure of AGN's power in our sample.

A simple least-square linear fit within a wide range of luminosity gives an empirical relationship
between $L_{[\mathrm{OIII}]}$ and $L_{\mathrm{H\alpha}}$
\begin{equation}
\log\left(\frac{L_{[\mathrm{OIII}]}}{10^{40}\ \mathrm{erg\ s^{-1}}}\right)=(1.518\pm0.06)+
(1.078\pm0.07)\log\left(\frac{L_{\mathrm{H\alpha}}}{10^{42}\ \mathrm{erg\ s^{-1}}}\right)
\end{equation}
Incorporating with Eq. 3 yields a relationship between continuum luminosity $L_{5100}$ and
$L_{\mathrm{[OIII]}}$: 
$L_{5100}/10^{42}\mathrm{erg\ s^{-1}}\approx1.5(L_{\mathrm{[OIII]}}/10^{40}\mathrm{erg\ s^{-1}})^{0.8}$.
Taking into account of the dynamical range of $L_{\mathrm{[OIII]}}$ from
$\sim10^{40}$ to $\sim10^{42}\ \rm{erg\ s^{-1}}$, 
an average ratio is $L_{5100}/L_{\mathrm{[OIII]}}\sim100$.
Although the ratio is three times lower than the mean value derived 
in Kauffmann et al. (2003a) and in Heckman et al. (2004), it 
is reasonable because of the large scatter in the correlation between $L_{\mathrm{[OIII]}}$
and $L_{5100}$. There are two possible
explanations for the discrepancy of the $L_{5100}/L_{\mathrm{[OIII]}}$ ratio: 1) The luminosity
of the broad H$\alpha$ line (and hence $L/L_{\mathrm{Edd}}$ and $M_{\mathrm{BH}}$) 
might be underestimated due to the obscuration caused by torus in the frame of
the unified model of AGN (Antonucci 1993).  It is noted that this systematic
shift would not seriously affect our final statistic trends, both because of the 
tight correlation between $L_{\mathrm{H\alpha}}$ and  $L_{\mathrm{[OIII]}}$ and 
because of the weak dependencies of $L/L_{\mathrm{Edd}}$ and $M_{\mathrm{BH}}$ on $L_{\mathrm{H\alpha}}$
(i.e., $L/L_{\mathrm{Edd}}\propto L_{\mathrm{H\alpha}}^{0.31}$ and 
$M_{\mathrm{BH}}\propto L_{\mathrm{H\alpha}}^{0.55}$, see Eqs. 3 and 4). 
2) In additional to the AGN's contribution, star formation can also contributes to the [\ion{O}{3}] luminosity.
In fact, our sample is mainly composed of the composites in which the narrow emission lines 
are likely to be mainly ionized by hot stars (e.g., Kewley et al. 2006, and see below).

We estimate the upper limits of the underestimates of both $L/L_{\mathrm{Edd}}$ and $M_{\mathrm{BH}}$ 
as follows. Assuming the underestimate of $L_{\mathrm{H\alpha}}$ is entirely
caused by the obscuration of torus and by the reddening in BLR, the observed broad H$\alpha$ luminosity
$L'_{\mathrm{H\alpha}}$ could be written as 
$L'_{\mathrm{H\alpha}}=f_{t}f_{e}L_{\mathrm{H\alpha}}$, where $L_{\mathrm{H\alpha}}$ is the intrinsic 
H$\alpha$ luminosity, $f_{t}$ and $f_{e}$ represents the decreases of the line luminosity due to the 
obscuration and reddening, respectively. If the three times lower in 
the $L_{5100}/L_{\mathrm{[OIII]}}$ ratio is caused by the obscuration, $f_t$ is inferred to be $\sim0.28$
according to the relation $L_{5100}\propto L_{\mathrm{H\alpha}}^{0.86}$.
The factor $f_e$ could be estimated from the 19 objects for 
which the broad H$\beta$ is present. Recent studies found the mean value of the broad-line Balmer decrements 
(H$\alpha$/H$\beta$) in typical type I AGNs is significantly close to the value of 3.1 
predicted by the case B recombination
(e.g., Greene \& Ho 2005; Vanden Berk et al. 2001; Dong et al. 2007). However, large
Balmer decrements (7-10) are observed in Seyfert 1.9-like galaxies (e.g., Osterbrock 1981).
Given the Balmer decrement in the case B recombination, 
only upper limit of $f_e$ could be obtained in our sample,
both because of the orientation effect and because of the fact that higher
density and/or higher ionization parameter is required to produce H$\beta$ than H$\alpha$. 
The mean value of the broad-line Balmer decrement
$\langle$H$\alpha$/H$\beta\rangle_\mathrm{B}$ is 5.97 in current sample.
Excluding the reddening inferred from 
the narrow-line Balmer decrement $\langle$H$\alpha$/H$\beta\rangle_\mathrm{N} =4.13$, 
we obtain $f_e\sim0.47$. 
Finally, the upper limit of the underestimate of $L/L_{\mathrm{Edd}}$ is estimated to be $\sim50$\%,
and that of $M_{\mathrm{BH}}$ to be $\sim70$\%.      

\subsection{Deriving age of stellar population of host}

Both 4000\AA\ break ($\rm D_n(4000)$) and equivalent width (EW) of the H$\delta$ absorption (H$\delta_A$)
are widely used in AGN's hosts as indicators of the ages of their stellar populations (e.g., Heckman et al. 2004; 
Kauffmann et al. 2003a; Kewley et al. 2006).  
Incorporating with the large high-resolution stellar spectra libraries, 
the latest theoretical population synthesis models of BC03 indicate that both $\rm D_n(4000)$ 
and H$\delta_{\rm{A}}$ are reliable age indicators until a few Gyrs after 
the burst (e.g., Figure 2 in Kauffmann et al. 2003c), 
although both indices are sensitive to metallicity in the very old stellar populations. 

The 4000\AA\ break is a strong discontinuity in the optical spectrum of a galaxy. 
The break is mainly caused by the absorption features of ionized metal lines blueward of 4000\AA. 
The popular definition of $\rm D_n(4000)$ is (Balogh et al. 1999; Bruzual 1983)
\begin{equation}
D_n(4000)=\frac{\int_{4000}^{4100}f_\lambda d\lambda}{\int_{3850}^{3950}f_\lambda d\lambda}
\end{equation}
Gorgas et al. (1999) calibrated $\rm D_n(4000)$ in terms of the stellar atmospheric parameters: effective
temperature, metallicity and surface gravity. In a young stellar population, 
small $\rm D_n(4000)$ arises from the 
low opacity of metal since it is multiply thermal ionized in hot stars. On the contrary, large 
$\rm D_n(4000)$ appears in an old stellar population in which the metal opacity is expected to be high.

The strength of the H$\delta$ absorption is an age indicator when the stellar populations are in the post-starburst phase, 
i.e. 0.1-1 Gyrs after the burst when massive stars (O and B stars) terminated their evolutions 
(Worthey \& Ottaviani 1997; Gonzalez Delgado et al. 1999). 
The index H$\delta_{\rm{A}}$ measures the EW of absorption in A type stars, and is defined
as (Worthey \& Ottaviani 1997)
\begin{equation}
\mathrm{H}\delta_A=(4122.25-4083.50)(1-\frac{F_I}{F_C})
\end{equation}
where $F_I$ is the flux within the $\lambda\lambda4083.50-4122.25$ feature bandpass, and $F_C$ the flux of
the pseudo-continuum within two defined bandpasses: blue $\lambda\lambda4041.60-4079.75$ and red
$\lambda\lambda4128.50-4161.00$. 
As the case of $\rm D_n(4000)$, similar calibration has been carried out by Worthey \& Ottaviani (1997).

We measure both indices in each starlight spectrum. 
The starlight spectrum is separated from the raw data as described in Section 2.2.
For each object, the measured values of $\rm D_n(4000)$ and H$\delta_{\rm{A}}$  
is listed in Column (7) and (8) in Table 1, respectively. 
The comparison between the values of $D_n(4000)$ obtained in this paper and 
that derived by MPA/JHU group is shown in the left panel in Figure 2. 
The same comparison but for H$\delta_A$ is shown in the right panel.
The left panel shows a significant correlation (with slope $\approx1$) 
between the two sets of $D_n(4000)$ values,
and the right panel a less significant, but still tight one for H$\delta_A$. The comparisons clearly
indicate that our measurements are highly consistent with those derived by MPA/JHU group, although
our H$\delta_A$ appears to be slightly smaller than those of MPA/JHU group at left-bottom corner.
This small discrepancy could be ignored in current study since 
we only focus on the statistic properties of the composite AGNs.
A large discrepancy is only  
found in one object SDSS\,J085338.27+033246.1. The value of H$\delta_A$ derived in this paper is 
7.22 which is much larger than H$\delta_A$=2.78 obtained by MPA/JHU group. 
The large discrepancy is due to the difference in the spectra components adopted to
model both continuum and absorption features. In addition to the seven eigenspectra, a
powerlaw continuum attributed to AGN is involved in the spectra modeling for the object,
which reduces its starlight continuum level, and enhances 
its measured H$\delta_A$ consequently. 

\section{Classification of the Composite AGNs}

In this section, we first examine the distributions of the physical properties derived in the last 
section on the traditional BPT diagrams, and then perform a very fine classification 
on the composite AGNs. Their host's properties are finally explored basing upon the classification. 

The BPT diagrams are used as a powerful tool to diagnose dominant energy source in
emission-line galaxies. Seyferts, LINERs, and star-forming galaxies can be separated according to
their emission-line ratios. With respect to star-forming galaxies, 
AGNs have larger ratios in [\ion{O}{3}]/H$\beta$, 
[\ion{N}{2}]/H$\alpha$, [\ion{S}{2}]/H$\alpha$, and [\ion{O}{1}]/H$\alpha$, 
because of their stronger and harder ionizing field.
AGNs therefore concentrate in the up-right corner in each BPT diagram. 
The theoretical curves discriminating between AGNs and star-forming 
galaxies in the BPT diagrams were recently proposed by Kewley et al. (2001).
There are many optical classification schemes to segregate LINERs from other emission-line galaxies
(e.g., Heckman et al. 1980, Veilleux \& Osterbrock 1987; Ho et al. 1997). In this study, we adopt the 
new empirical classification scheme recently proposed by Kewley et al. (2006). 
The classification scheme was obtained by analyzing a large sample
of emission-line galaxies selected from SDSS survey. These new classification line is very 
similar to that of Heckman et al (1980) in both slope and intersection.

The BPT diagrams are displayed in Figure 3 for the composite AGNs. 
The three panels correspond to the three different methods for classifying 
galaxies using two pairs of line ratios. In each
panel, the theoretical classification line is shown by a solid line,
and the empirical classification line suggested by Kauffmann et al. (2003a) by a dashed line. 
The classification lines used to separate Seyferts from LINERs are shown by dot-dashed lines. 
The size of each point is scaled to be proportional to the corresponding value of $D_n(4000)$ simply 
because of the high level of accuracy of this parameter (See Section 3.2 for details). 

At first, almost 
all the composite AGNs are located between the theoretical and empirical lines in the 
[\ion{O}{3}]/H$\beta$ vs. [\ion{N}{2}]/H$\alpha$ diagram, which indicates that
our AGN-host separations and emission-line measurements are consistent with those of MPA/JHU groups.
In addition, it is interesting that 
the objects occupy a narrow vertical stripe with nearly constant line ratio 
[\ion{N}{2}]/H$\alpha$. The examination of this phenomenon is out of the scope of this paper, and 
will be studied in subsequent works. Moreover, no clear trend is found along this
stripe for $D_n(4000)$, although the objects with larger $D_n(4000)$ are only found at the top end of the stripe. 
In contrast, clear trends are found in the other two panels.
The objects located in the occupations of LINERs generally have larger 
$D_n(4000)$ with respect to that located below the theoretical demarcation lines,
which implies that the composite AGNs at least could be divided into two sub-groups:
\ion{H}{2} region-like ionization + young stellar population and LINER-like ionization + old stellar population  
\footnote{Similar phenomena and indications could be found in the 
case of $L/L_{\mathrm{Edd}}$ and luminosity of [\ion{O}{3}] instead of $D_n(4000)$, with, 
however, less significance.}. 

Since the composite AGNs are initially selected only from the [\ion{N}{2}]/H$\alpha$ vs. [\ion{O}{3}]/H$\beta$ diagram, 
a more fine classification on the objects is performed as similar as done in Kewley et al. (2006). 
We separate our sample into three groups according to 
the classification scheme described as follows: 
a) The \ion{H}{2} composite AGNs are galaxies that lie below the theoretical demarcation curves
in all the [\ion{N}{2}]/H$\alpha$, [\ion{S}{2}]/H$\alpha$ and 
[\ion{O}{1}]/H$\alpha$ diagrams. These galaxies correspond to the ``composite galaxies'' in Kewley et al. (2006); 
b) In contrast, the Seyfert+LINERs are galaxies that lie above the theoretical demarcation 
curves in all the three diagrams; c) The remaining are multiply-classified galaxies that are classified
differently in the three diagrams. These galaxies correspond to the case b ``ambiguous galaxies'' in
Kewley et al. (2006).
The classification is shown in the three BPT diagrams in Figure 4. The demarcation lines are
shown in the figure by the same symbols used in Figure 3.
Only 74 objects for which all the line ratios are
determined are plotted. The \ion{H}{2} composite AGNs, Seyfert+LINERs and multiply-classified
galaxies are shown by red, blue, and black points, respectively. 

Our detailed classification indicates
that, among the 74 objects, six ones can be classified as Seyfert+LINERs rather than composites.
The fraction of these six objects is so small that their existence has negligible effect on 
the overall properties of our sample. Totally,
there are 45 \ion{H}{2} composite AGNs and 23 multiply-classified galaxies. The comparison in
the age of stellar population between  
the \ion{H}{2} composite AGNs and multiply-classified galaxies is illustrated in Figure 5.
The dashed line shows distribution of $D_n(4000)$ for the \ion{H}{2} composite AGNs, and 
the solid line for the multiply-classified galaxies.
A narrow distribution with median value $D_n(4000)$=1.30 and standard deviation of 0.11
is found in the \ion{H}{2} composite AGNs.
On the contrary, a very wide and possibly 
bimodal distribution is identified in the multiply-classified galaxies.

\section{Star Formation History and Evolution of AGN}

We plot our sample of the composite AGNs in the $D_n(4000)$/H$\delta_A$ plane in Figure 6.
The dashed line shows the evolution of the standard SSP model with solar metallicity, and the dot-dashed line
the evolution of the model with continuous star formation history (i.e., an exponentially declining SFR
$\psi(t)\propto e^{-t/(4\rm Gyr)}$) at solar metallicity (BC03). The $D_n(4000)$/H$\delta_A$
plane is a useful diagnosis to determine whether a galaxy has been forming stars continuously or
in a recent burst. Galaxies with continuous star formation history occupy a narrow locus
in the plane. A recent burst ended in 0.1-1 Gyrs ago results in a significant 
displacement away from the locus (i.e., be larger in H$\delta_A$), because the optical spectrum 
is dominated by the emission of A type stars. Even though the small discrepancy in 
H$\delta_A$ is found between our measures and those in MPA/JHU groups, the composite AGNs
fall close to the model with continuous star formation history, except the 
object SDSS\,J085338.27+033246.1 with significant strong H$\delta$ absorption H$\delta_A$=7.22 (see 
Section 3.2 for explanation).

Our sample allows us to investigate the issue of co-evolution of AGN and star formation 
in a new approach. Our study differs from the previous investigations in: 1)
it should be emphasized that each composite AGN
certainly contains a Seyfert 1 nucleus in its center, because of the existence of the broad H$\alpha$
emission. In fact, there might be two populations of Seyfert 2 galaxies: with and without a hidden
Seyfert 1 nucleus. The spectropolarimetric surveys and X-ray observations indicate that the
hidden Seyfert 1 nuclei are indeed not present in a fraction (10\%-50\%) of Seyfert 2 galaxies
(e.g., Tran 2001, 2003; Moran et al. 2000, 2001, 2007; Panessa \& Bassini 2002; Gallo et al. 2006;
Pappa et al. 2001); 2) In addition, the basic properties of AGN (i.e., $L/L_{\mathrm{Edd}}$ and $\rm M_{BH}$)
are directly obtained from the
nuclear emission rather than the tracers with large uncertainties. For instance,
the luminosity of [\ion{O}{3}] emission line is statistically used as a tracer of AGN's bolometric 
luminosity in Seyfert 2 galaxies by many authors (e.g., Kauffmann et al. 2003a; Heckman et al. 2004; Kewley et al. 2006).

It was suspected that $L/L_{\mathrm{Edd}}$ might be of significance of ``age'' of AGN 
(e.g., Grupe et al. 1999; Grupe 2004; Wang et al. 2006). In order to examine this point, the
$L/L_{\mathrm{Edd}}$ vs. $D_n(4000)$ diagram is presented in the \sl Panel A \rm in Figure 7.
The diagram indicates that the $L/L_{\mathrm{Edd}}$ estimated from the broad H$\alpha$ 
emission varies as a function of the age of the host's stellar population.
The black holes with high 
specific accretion rates tend to reside in the hosts with relatively young stellar populations.
This dependence therefore strongly supports the evolutionary
scenario of AGN in which AGNs with high $L/L_{\mathrm{Edd}}$ evolve to ones with low $L/L_{\mathrm{Edd}}$ 
(Grupe 2004; Wang et al. 2006; Wang \& Wei 2006). 
The \sl Panel C \rm in Figure 7 shows a variation of the $L/L_{\mathrm{Edd}}$ with the line ratio 
[\ion{O}{1}]/H$\alpha$ which is sensitive to the hardness of the ionizing field. As shown in 
the diagram, a soft ionizing spectrum tends to be associated with a relatively high 
$L/L_{\mathrm{Edd}}$ on average, and a hard spectrum to be associated with an extremely low $L/L_{\mathrm{Edd}}$ only. 
This trend is consistent with the strong correlation identified between the hardness of hard/soft X-ray emission 
and $L/L_{\mathrm{Edd}}$ for both local and distant AGNs (Grupe 2004; Bian 2005; Shemmer et al. 2006). 
The lowest value of the $L/L_{\mathrm{Edd}}$ among our sample is 0.003, which is consistent with 
the prediction of the advection dominated accretion flow (ADAF) model. 
The model can only produce sub-Eddington accretions ($L/L_{\mathrm{Edd}}\leq0.01-0.1$).
Because a large fraction of the energy released is advected 
with hot gas into black hole rather than radiated by emission, the gas is extremely hot, and 
produces a harder spectrum in a low state 
(e.g., Abramowicz et al. 1995; Narayan \& Yi 1995; Esin et al. 1997; Meyer et al. 2000; 
Ferreira et al. 2006).

The another candidate of ``age'' of AGN is the accretion rate onto a black hole
(i.e., luminosity released by accretion). Relatively good relations are identified between 
$L_{\mathrm H\alpha}$ and $D_n(4000)$, and between $L_{\mathrm H\alpha}$ and [\ion{O}{1}]/H$\alpha$. 
The two relations are presented in Figure 8.
Noted that both relations degrade at the right-bottom end. This phenomenon does not appear 
when $L/L_{\mathrm{Edd}}$ is used in place of $L_{\mathrm H\alpha}$ as shown in the \sl Panel A \rm 
and \sl C \rm in Figure 7, which indicates that the $L/L_{\mathrm{Edd}}$ is the major physical driver
in the evolution of AGN.

A significant correlation between $D_n(4000)$ and the line ratio [\ion{O}{1}]/H$\alpha$ is first shown in 
the \sl Panel B \rm in figure 7. The tight correlation implies that AGN evolves from the high state 
with a soft spectrum to the low state with a hard spectrum. The line ratio [\ion{O}{1}]/H$\alpha$ is 
suggested to be used as a potential indicator of the age of stellar population.

\section{Discussion}

In addition to the AGN's contribution, other main mechanisms of the broad Balmer emission lines 
are the stellar envelopes of Wolf-Rayet stars and OB associations associated multiple SN events 
(Izotov et al. 2007 and references therein). Both mechanisms can produce the broad emission lines
with gaseous velocity as large as $\stackrel{>}{\scriptstyle\sim} 1000\ \mathrm{km\ s^{-1}}$. 
The luminosity of the broad H$\alpha$ emission line ranges from $10^{40}$ to $10^{42}\ \mathrm{ergs\ s^{-1}}$
in the composite AGNs, which greatly exceeds the typical luminosity of stellar winds from WR or LBV stars 
($10^{36}$-$10^{40}\  \mathrm{ergs\ s^{-1}}$). Moreover, the broad H$\alpha$ emission in 
the composite AGNs is unlikely caused by the SN events, because of the existence of the  
underlying relatively old stellar populations in the spectra.

\subsection{Aperture effect}
The redshift ranges roughly from 0.05 to 0.15 in our sample.
In this range of distance, the typical projected fiber diameter ($3$\symbol{125}) is from $\sim3$kpc to $\sim8$kpc.
When a spectrum is taken in the SDSS spectroscopic survey, the fixed fiber aperture
would result in more contamination of the light emitted from the outer part of host galaxy in
distant galaxies than in nearby galaxies. 
First, the effect leads the LINER fraction decreases with redshift because the weak 
emission lines are more difficult to be detected in distant galaxies than in nearby galaxies. 
Such effect exists in our sample, and is shown in the left panel in Figure 9. 
The fraction of the LINER-like galaxies with an old stellar population drops significantly when redshift 
$z>0.1$, which is consistent with Kauffmann et al. (2003a) and Kewley et al. (2006). 
Secondly, for a fixed aperture, as the contamination increases 
with redshift, the value of $D_n(4000)$ would increase with redshift if a circumnuclear starburst occurs.
If so, one could see a lower boundary with positive slope in the $D_n(4000)$ vs. z diagram, 
which is, however, not present in the diagram. The aperture effect can not explain why no 
young stellar populations ($D_{4000}\leq1.2-1.4$) is deteced in the low-redshift galaxies. 
The negative trend shown in the diagram 
could be explained by the consequence of the increase of $L_{\mathrm{H\alpha}}$ with the redshift.
Moreover, the lower limit of redshift of $z\sim0.05$ ensures that the spectra
taken within $3$\symbol{125} aperture could approximately reflect the global properties (Kewley et al. 2005).
The right panel in Figure 9 shows the broad-to-narrow H$\alpha$ line ratio as a function of the redshift. 
As shown by the diagram, in general, our criterion for the existence of the broad H$\alpha$ line 
is free of the aperture effect.

\subsection{Star formation rate}
In section 4, we have built a sub-sample of the \ion{H}{2} composite AGNs which
are below the theoretical demarcation lines in the three BPT diagrams. 
The left panel in Figure 10
shows the [\ion{O}{3}]/[\ion{O}{2}] vs. [\ion{O}{1}]/H$\alpha$ diagram for the
\ion{H}{2} composite AGNs.
The distribution shows that the \ion{H}{2} composite AGNs
lie within the occupation of the star-forming galaxies, which
additionally suggests that hot stars might be main contributor of the ionizing field for the narrow emission lines.
The luminosity of [\ion{O}{2}] emission line is
an indicator of current star formation rate (SFR) for \ion{H}{2} regions as suggested by
Kennicutt (1998) and Kewley et al. (2004). However, the origin of the [\ion{O}{2}] emission is complex in the 
\ion{H}{2} composite
AGNs, since both AGN and starburst can contribute to [\ion{O}{2}] (e.g., Kim et al. 2006; Yan et al. 2006).
In order to estimate the contribution by AGN, the right panel in Figure 10 shows the luminosity of [\ion{O}{3}]
as a function of the line ration [\ion{O}{2}]/[\ion{O}{3}]. The solid line is the least-squares bisector
regression for Type I AGNs given by Kim et al. (2006). The \ion{H}{2} composite AGNs clearly show enhanced 
[\ion{O}{2}]/[\ion{O}{3}] ratios. Similar as done in Kim et al. (2006), 
we estimate the [\ion{O}{2}] luminosities emitted from \ion{H}{2} region by 
assuming the enhanced [\ion{O}{2}]/[\ion{O}{3}] ratios are caused by star formations. 
The mean value of the [\ion{O}{2}]/[\ion{O}{3}] ratio is $\sim1.0$ for the \ion{H}{2} composite AGNs. 
Given the average [\ion{O}{3}] luminosity $L_{[\mathrm{OIII]}}\sim3\times10^{41} \mathrm{\ erg\ s^{-1}}$, 
the [\ion{O}{2}]/[\ion{O}{3}] ratio predicted by the regression is only $\sim0.25$. Then, on average, 
$\sim$75 percent of the [\ion{O}{2}] emission is estimated to come from star formation in the \ion{H}{2} composite AGNs.

We use the recent calibration (Kewley et al. 2004)
\begin{equation}
\mathrm{SFR}=7.9\times\frac{L_{\mathrm{[OII],42}}}{16.73-1.75[\log(\mathrm{O/H})+12]}M_{\odot}\rm{yr^{-1}}
\end{equation}
to estimate current SFRs, where $L_\mathrm{[OII],42}$ is the luminosity of [OII]
emission in units of $10^{42}\ \rm{ergs\ s^{-1}}$. The metallicity is fixed to be $\log (\rm{O/H})+12=9.2$,
i.e., twice of the solar, as suggested by Ho (2005b).

A correlation between the current SFRs and $L/L_{\mathrm{Edd}}$ is
shown in the \sl left \rm panel in Figure 11. The relatively large scatter could be caused by 
both the underestimation of $L_{\mathrm{H\alpha}}$ and the contribution of AGN to $L_{\mathrm{[OII]}}$. 
The correlation indicates that a more powerful AGN tends to be
associated with a more intense current star formation. The relation is reasonable because
the strengths of both AGN and star-forming activities depend on the consumption of gas.  
Heckman et al. (2004) found a roughly constant ratio of the star formation rate to the accretion rate onto black 
hole in the SDSS survey. The dependence of the SFR on the
luminosity of AGN is further supported by the correlation,
between the X-ray luminosity in 0.5-4.5 keV band and CO luminosity, found in Seyfert 1 galaxies
and in quasars (Yamada 1994).
The \sl middle \rm and \sl right \rm panels
in Figure 11 shows variations of the SFRs with $D_n(4000)$ and H$\delta_A$, respectively.
As shown by the diagrams, the current SFR decreases when stellar populations passively evolves,
because of the depletion of the gas. Both relations are consistent with Figure 6 in which a continuous star
formation history with an exponentially decaying SFR is required to match the observations to the modeled evolution locus.
In addition, the [\ion{O}{2}] inferred SFRs are typically below 10$\rm{M_{\odot}yr^{-1}}$.
The suppressed current SFRs are identified in a number of luminous AGNs by several authors (e.g., Ho 2005b;
Kim et al. 2006; Wang \& Wei 2006; Martin et al. 2007).

\subsection{Role of $L/L_{\mathrm{Edd}}$ in AGN evolution}

Similar as the scenario proposed by Weedman (1983),
Wang \& Wei (2006) recently proposed a scenario of co-evolution of AGN and star formation in its host
basing upon the differential growth of the black hole mass and bulge mass.
In the scenario, after a starburst occurs first and triggers accretion onto the central black hole latter, 
AGN then shines with decreasing $L/L_{\mathrm{Edd}}$ and insignificant increasing black hole mass, 
as the young stellar population ages and fades. The scenario is supported
by the prediction of the model developed by Kawakatu et al. (2003). The model predicts that the AGN-dominated
phase follows the host-dominated phase in a time scale of a few $10^8$ years. This change in phase is 
also predicted in the more solid physical models developed by Granato et al. (2001, 2004).

In addition to the models, there were accumulating observations supporting 
that $L/L_{\mathrm{Edd}}$
plays an important role in the evolution of galaxy in recent years. Mathur (2000) suggested
that NLS1 might be young AGN with high $L/L_{\mathrm{Edd}}$ and small black hole mass. The young stellar
populations within post-starburst phase are found to be associated with NLS1s in a few cases 
(e.g., Wang et al. 2004; Wang \& Wei 2006; Zhou et al. 2005). 
The evolutionary significance of $L/L_{\mathrm{Edd}}$
was suspected by Grupe (2004). In two years latter, 
Wang et al. (2006) extended the well-documented E1 space into the infrared color
$\alpha(60,25)$ that is an indicator of relative strength of star formation with respect to AGN. 
The infrared-color dominated E1 space 
implies a relation between the $L/L_{\mathrm{Edd}}$ and star formation history. With the advent of
SDSS, Kewley et al. (2006) and Wild et al. (2007, and references therein) revealed a relationship
between the star formation history and $L_{\mathrm{[OIII]}}/\sigma_*^4$ in type II AGNs . 

Taking advantage of the composite AGNs, 
the evolutionary significance of $L/L_{\mathrm{Edd}}$ (or the E1 space) is obviously
demonstrated by the clear dependence of $L/L_{\mathrm{Edd}}$ on $D_n(4000)$. 
In addition, the evolutionary significance of the E1 space is also implied by the tight 
correlation between [\ion{O}{1}]/H$\alpha$ and $D_n(4000)$. The line ratio [\ion{O}{1}]/H$\alpha$ is known 
to be sensitive to the hardness of the ionizing radiation field. The X-ray photon index is an important 
component in the E1 space 
(e.g., Wang et al. 1996; Sulentic et al. 2000; Xu et al. 2003; Grupe et al. 1999; Grupe 2004).
In summary, by adopting the new approach, our results are not only consistent with the previous studies, but also
strongly support the argument that AGN evolves along the E1 sequence from 
high $L/L_{\mathrm{Edd}}$ end to low 
$L/L_{\mathrm{Edd}}$ end as stars evolve from young to old stellar population.

\subsection{``Young'' and ``Old'' composite AGNs}

Our results indicate that $\sim$80\% of the composite AGNs are high $L/L_{\mathrm{Edd}}$
emitters associated with a post-starburst (with age less than
1\,Gyr, corresponding to a value $D_n(4000)\leq1.5$).
In the context of the evolutionary scenario, the composite AGNs associated with both
high $L/L_{\mathrm{Edd}}$ and young stellar population might be ``Young'' AGNs.  
The spatial connection between the post-starburst and AGN is supported by the spatial 
resolved spectroscopy in the three AGNs selected from SDSS (Goto 2006), and by recent photometric observations 
with spatial resolution of 0.1-0.2\symbol{125} (Davies et al. 2007). In terms of the equivalent width 
of the narrow H$\alpha$ emission, Westoby et al. (2007) found the composite galaxies peak in the valley 
between the red and blue sequence (e.g., Strateva et al. 2001; Baldry et al. 2006), which suggests 
the composite galaxies are in the transition state. By calculating a series of photoionization models,
Stasinska et al. (2006) suggested that the right wing in the BPT
diagram could be explained by the variation of the balance between massive stars and AGN powers along the wing.
In the BPT diagram, a galaxy moves toward right-up direction from the occupation of 
the composite AGNs as the fraction of AGN's power increases.

On the other hand, although the multiply-classified galaxies mostly lie
below the theoretical demarcation curve in the [\ion{N}{2}]/H$\alpha$ vs. [\ion{O}{3}]/H$\beta$ diagram,
a fraction of them are located in the LINER region in the [\ion{O}{1}]/H$\alpha$ vs. [\ion{O}{3}]/H$\beta$ diagram.
The harder ionizing field and smaller AGN's power could be inferred in these multiply-classified galaxies,
because of the saturation of [\ion{N}{2}]/H$\alpha$ with increasing nebular metallicity in \ion{H}{2} region 
(e.g., Kewley \& Dopita 2002; Pettini \& Pagel 2004). We suspect that the multiply-classified galaxies 
associated with very old stellar populations are at the end of life of AGNs.

\section{Summary}
The goal of this paper is to study the evolutionary connection between AGN and its host galaxy in 
partially obscured AGNs (i.e., Seyfert 1.8 and 1.9 galaxies). Not as the previous studies,
we can derive the properties of both AGN and its host together in these objects
by direct measurements rather than some tracers. 
Taking advantage of the SDSS spectroscopic survey with large number of spectra of 
emission-line galaxies, 
we searched for Seyfert 1.8 and 1.9 galaxies (composite AGNs) from the emission-line 
galaxies (MPA/JHU catalogs), which are located between the empirical and theoretical classification lines
in the [\ion{O}{3}]/H$\beta$ vs. [\ion{N}{2}]/H$\alpha$ diagram.
The analysis of these 85 composite AGNs allows us to draw results as follows: 

\begin{enumerate}

\item The distribution of the composite AGNs on the [\ion{N}{2}]/H$\alpha$ 
vs. [\ion{O}{3}]/H$\beta$ diagram shows a 
narrow vertical stripe with nearly constant line ratio [\ion{N}{2}]/H$\alpha$. 
In contrast, the composite AGNs can be separated in the other two BPT diagrams, in which the 
objects associated with the very old stellar populations are located in the occupation of LINERs.

\item The BPT diagrams allow us to further divide the 68 composite AGNs into two subgroups:
45 HII composite AGNs which are below the theoretical
separation lines in all the three
BPT diagrams; 23 multiply-classified galaxies which are classified differently
in the three BPT diagrams. All the HII composite AGNs
are associated with the young stellar populations ($<$1Gyrs), 
while a very wide and possibly bimodal distribution of 
the age of the stellar population is identified in  
the multiply-classified galaxies. Totally, $\sim$80\% of the composite AGNs
are associated the young stellar populations within post-starburst phase ($D_n(4000)\leq1.5$). We suspect that 
the composite AGNs associated with the very old stellar populations are at the end of life of AGNs.

\item The broad H$\alpha$ emission inferred $L/L_{\mathrm{Edd}}$ forms a smooth sequence
with the age of the stellar population of AGN's host as assessed by $D_n(4000)$, 
which confirms the previous studies using other
tracers (e.g., $L_{\mathrm{[OIII]}}/\sigma_*^4$, infrared color $\alpha(60,25)$). The 
sequence degrades when $L/L_{\mathrm{Edd}}$ is replaced by $L_{\mathrm{H\alpha}}$. 
In addition, the $L/L_{\mathrm{Edd}}$ roughly depends on the line ratio
[\ion{O}{1}]/H$\alpha$ which is sensitive to the hardness of the ionizing field. 
Both theoretical population synthesis models and
[\ion{O}{2}] inferred SFR $\sim1-10M_{\odot}\mathrm{yr^{-1}}$ show a suppressed
current star formation in the composite AGNs. These results strongly support the evolutionary 
scenario suggested in Wang \& Wei (2006).

\item We first find a significant correlation between the line ratio [\ion{O}{1}]/H$\alpha$ 
and $D_n(4000)$ in the composite AGNs. If it is confirmed, the 
line ratio could be used to trace the age of stellar population in type I AGNs.

\end{enumerate}




\acknowledgments

We would like to thank the anonymous referee for his/her very useful comments and suggestions.
We thank D. W. Xu, C. N. Hao, Y. C. Liang and C. Cao for valuable discussions. We are
grateful to Todd A. Boroson and Richard F. Green for providing us with the 
\ion{Fe}{2} template. This work was supported by the National Science Foundation 
of China (grants 10503005 and 10473013). The SDSS achieve data are created and 
distributed by the Alfred P. Sloan Foundation.

\clearpage



\begin{figure}
\includegraphics[width = 13cm]{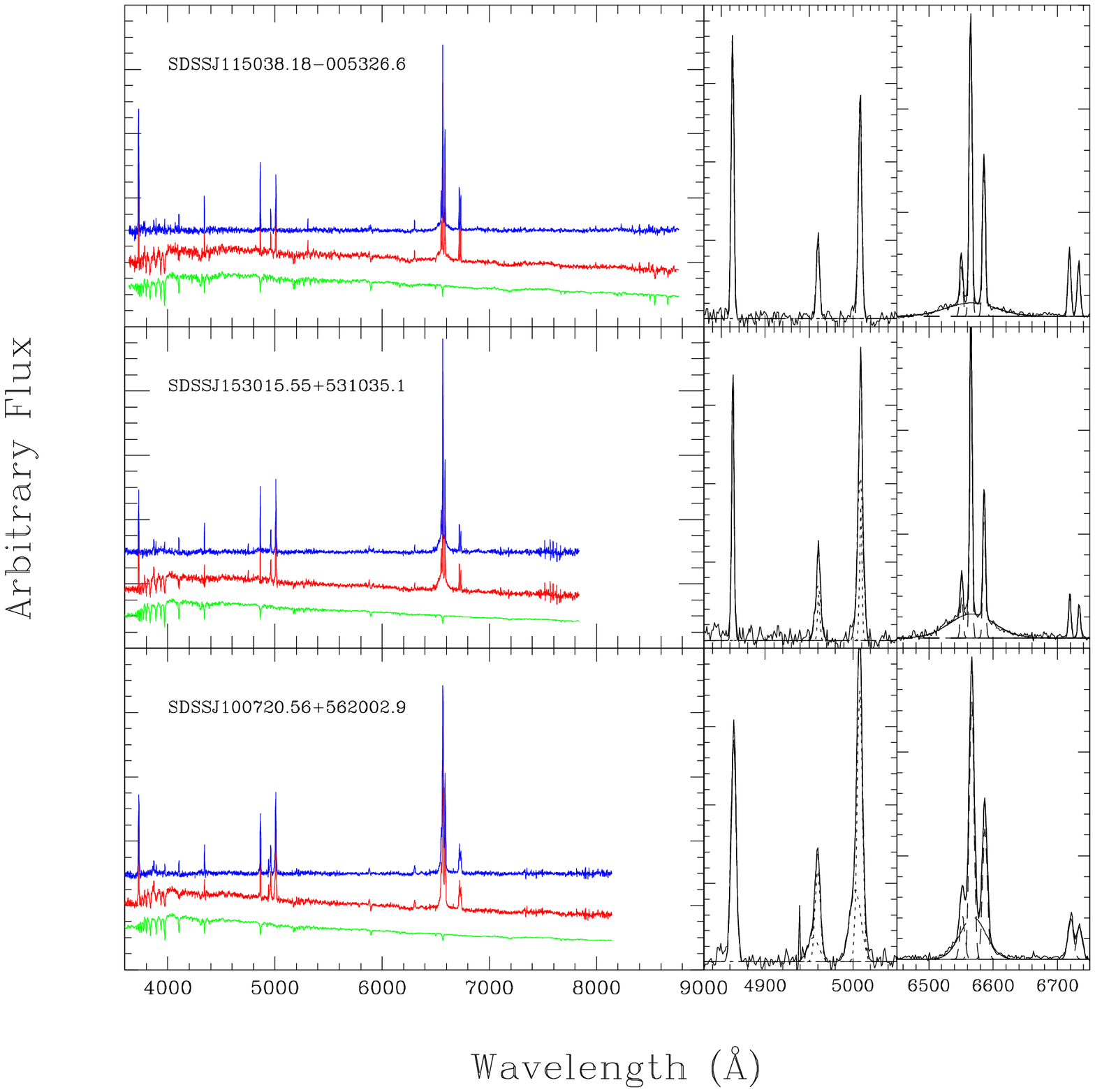}
\caption{\it Left column: \rm the modelings and subtractions of the starlight components by the
linear combination of the seven reddened eigenspectra in the three typical cases. In each panel, 
we plot the emission-line spectrum, observed spectrum and modeled starlight spectrum
from top to bottom. The spectra are vertically shifted by arbitrary amounts for visibility;
\it Middle column: \rm the modelings of the emission-line profile by a set of Gaussian components 
for H$\beta$ region; \it Right column: \rm the same as the middle column but for H$\alpha$ region. 
[\it see the electronic edition of the Journal for a color version of this figure.\rm]}
\end{figure}

\clearpage


\begin{figure}
\includegraphics[width = 13cm]{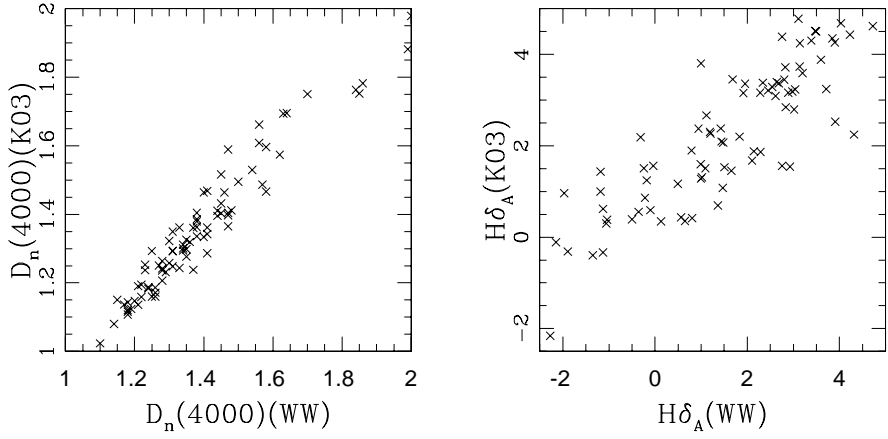}
\caption{The comparisons between the value obtained in this paper and those derived by MPA/JHU groups
for both indices $D_n(4000)$ (left panel) and H$\delta_A$ (right panel).
It is clear that the two independent measurements are highly consistent with each other. 
}
\end{figure}

\begin{figure}
\includegraphics[width = 12cm]{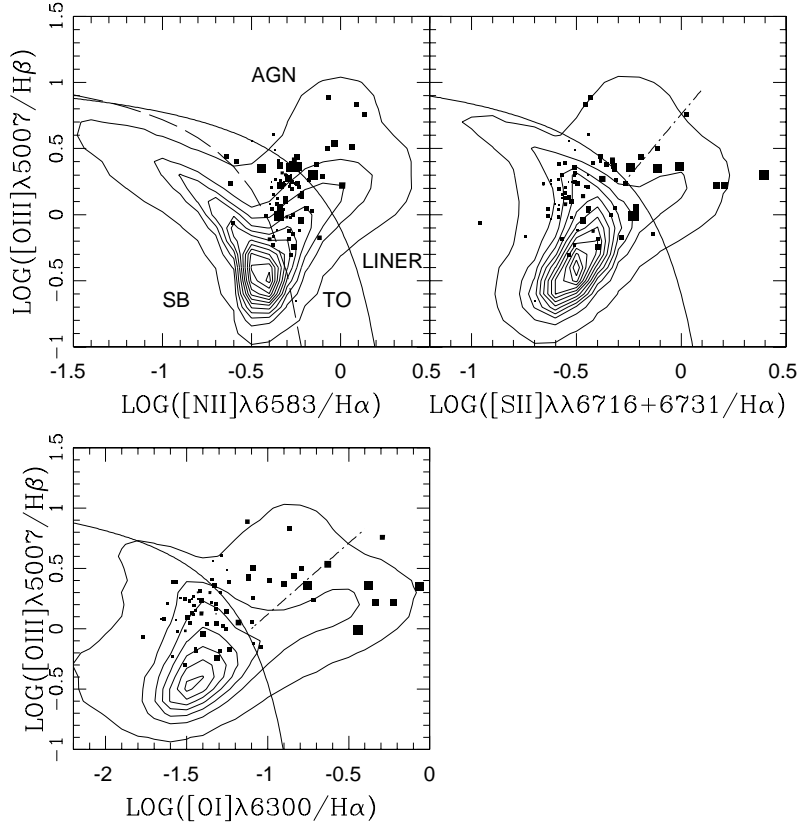}
\caption{The three diagnostic BPT diagrams for the composite AGNs. The theoretic demarcation 
lines separating AGNs from star-forming galaxies proposed by Kewley et al. (2001) are shown by the 
solid lines, and the empirical line proposed by Kauffmann et al. (2003a) by the dashed line. The dot-dashed 
lines drawn in the [\ion{S}{2}]/H$\alpha$ vs. [\ion{O}{3}]/H$\beta$ and [\ion{O}{1}]/H$\alpha$ vs. 
[\ion{O}{3}]/H$\beta$ diagrams show the empirical separation scheme of LINERs 
proposed in Kewley et al. (2006). The size of each point is proportional to the corresponding 
value of $D_n(4000)$. In the [\ion{N}{2}]/H$\alpha$ vs. [\ion{O}{3}]/H$\beta$ diagram, 
the composite AGNs distribute in a narrow vertical stripe with nearly 
constant line ratio [\ion{N}{2}]/H$\alpha$. However, the composite AGNs with very old stellar
populations (large $D_n(40000$) are mostly located in the LINER region in both 
[\ion{S}{2}]/H$\alpha$ vs. [\ion{O}{3}]/H$\beta$ and [\ion{O}{1}]/H$\alpha$ vs.
[\ion{O}{3}]/H$\beta$ diagrams.
}
\end{figure}

\begin{figure}
\includegraphics[width = 13cm]{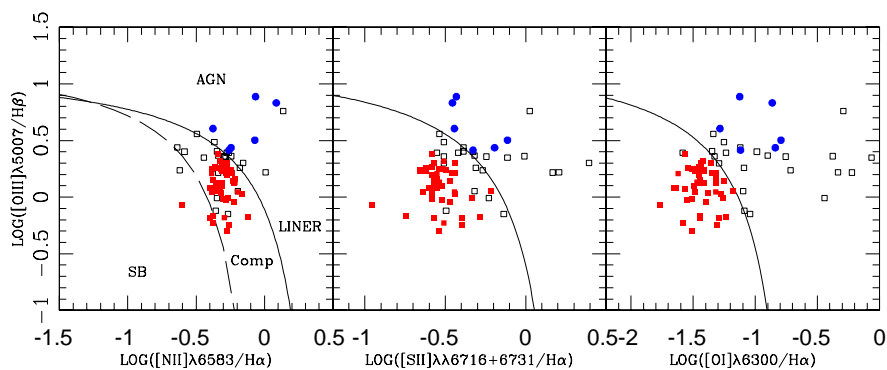}
\caption{The three diagnostic BPT diagrams showing our fine classification on the 74 composite AGNs. 
The displayed curves are the same as in Figure 3. 
The HII composite AGNs located below the theoretic lines 
in all the three BPT diagrams are presented by the red solid squares, and the Seyferts+ LINERs located
above the lines by the blue circles. The black open squares represent the multiply-classified galaxies 
which are classified differently in the three diagrams. [\it see the electronic edition of the Journal 
for a color version of this figure.\rm]} 
\end{figure}

\begin{figure}
\includegraphics[width = 13cm]{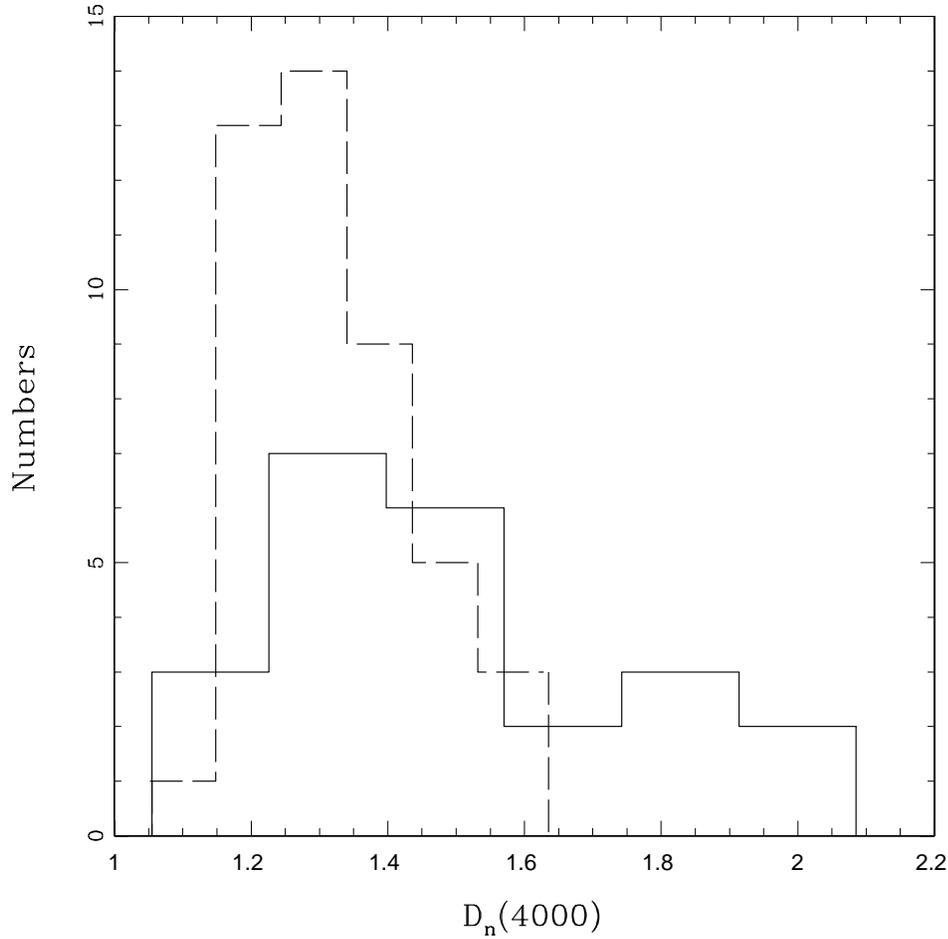}
\caption{The histograms of the index $D_n(4000)$ for the HII composite AGNs (the dashed line) 
and for the multiply-classified galaxies (the solid line). 
All the HII composite AGNs are associated with young stellar populations ($D_n(4000)<1.6$), while
the multiply-classified galaxies
show very broad and possibly bimodal distribution.}
\end{figure}

\begin{figure}
\includegraphics[width = 13cm]{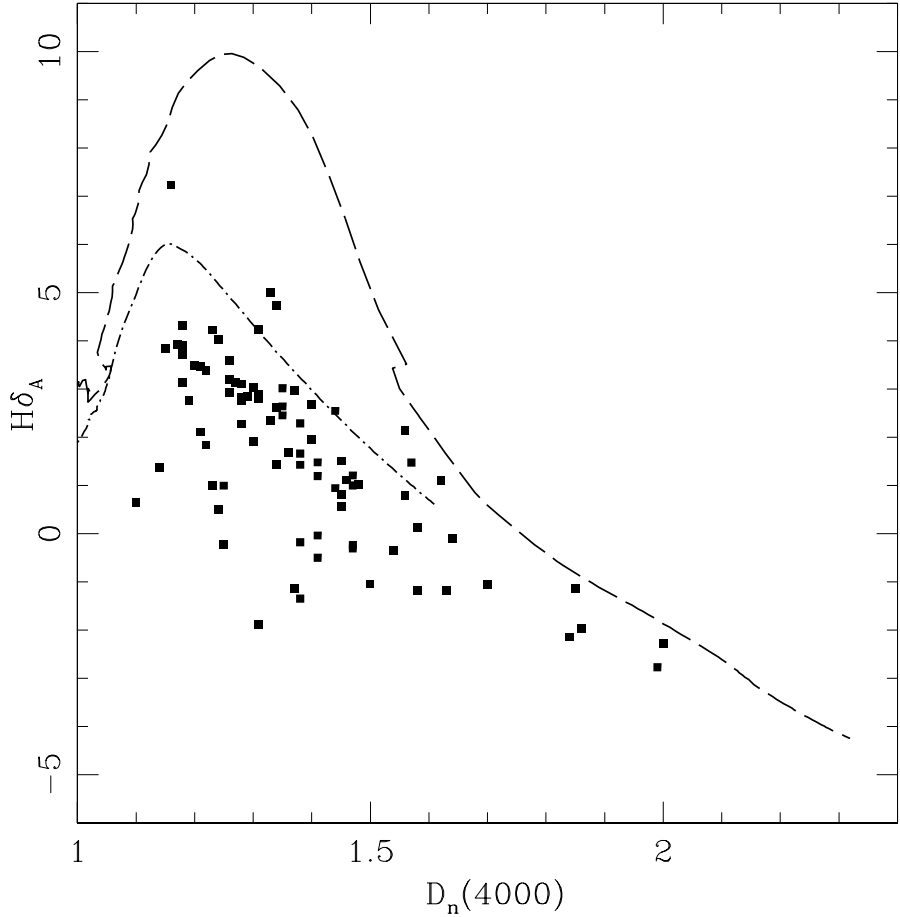}
\caption{The $D_n(4000)$/H$\delta_A$ plane for the composite AGNs. 
The dashed line represents the stellar population 
evolution locus of the SSP model with solar metallicity, and the dot-dashed line the model with exponentially 
decreasing star formation rate $\psi(t)\propto e^{-t/(\mathrm{4Gy})}$. The 
model with continuous star formation history is required to explain the composite AGNs.
} 
\end{figure}

\begin{figure}
\includegraphics[width = 13cm]{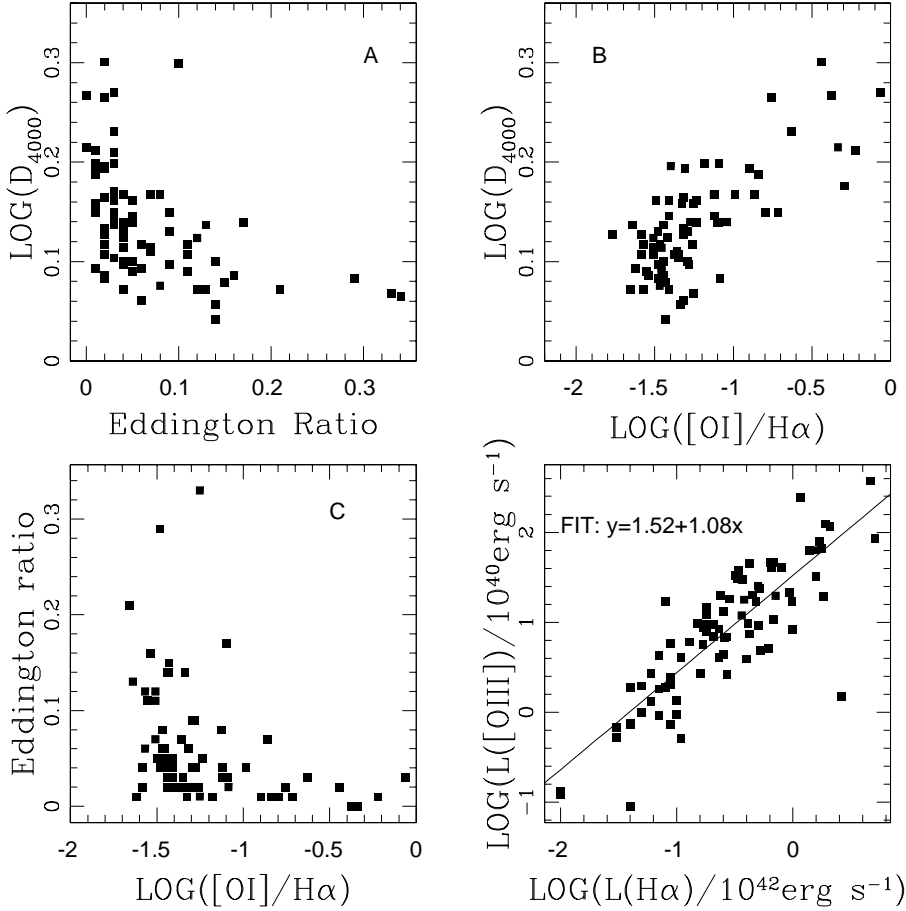}
\caption{\it Panel A: \rm A sequence of $L/L_{\mathrm{Edd}}$ with the age of 
stellar population of host for the composite AGNs. 
\it Panel B: \rm A tight correlation between $D_n(4000)$ and 
the line ratio [\ion{O}{1}]/H$\alpha$.
\it Panel C: \rm $L/L_{\mathrm{Edd}}$ plotted against the ratio [\ion{O}{1}]/H$\alpha$. 
As shown by the plot, there seems to 
be a trend that larger [\ion{O}{1}]/H$\alpha$ is statistically associated with lower 
$L/L_{\mathrm{Edd}}$. \it Panel D: \rm A tight correlation between the luminosity of 
the H$\alpha$ broad component and that of [\ion{O}{3}]. The best unweighted fitting is 
represented by the solid line.
}  
\end{figure}

\begin{figure}
\includegraphics[width = 13cm]{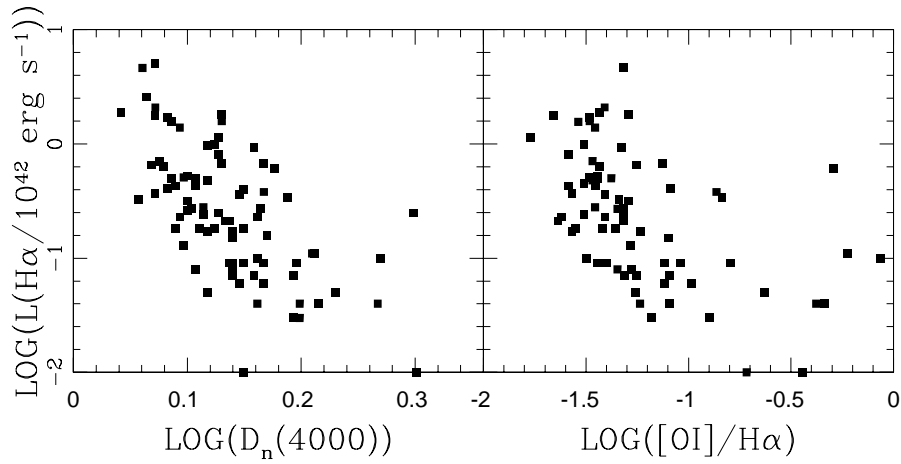}
\caption{
The luminosity of the H$\alpha$ broad component plotted against 
$D_n(4000)$ (\it left panel\rm) and H$\delta_A$ (\it right panel\rm). Although $L_{\mathrm H\alpha}$
is related with both $D_n(4000)$ and H$\delta_A$ at the high luminosity end,
both correlations degrade at the low luminosity end.}
\end{figure}

\begin{figure}
\includegraphics[width = 13cm]{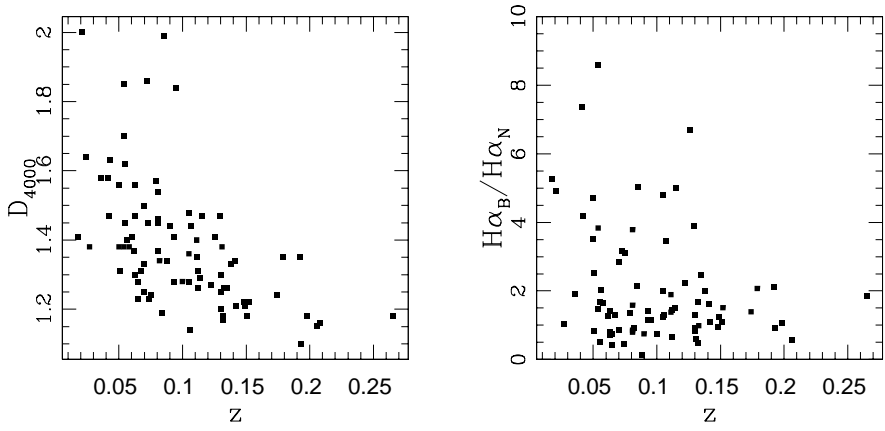}
\caption{The parameter $D_n(4000)$ (\it left panel\rm) and broad-to-narrow H$\alpha$ line ratio  
(\it right panel\rm) as a function of redshift. 
}
\end{figure}


\begin{figure}
\includegraphics[width = 13cm]{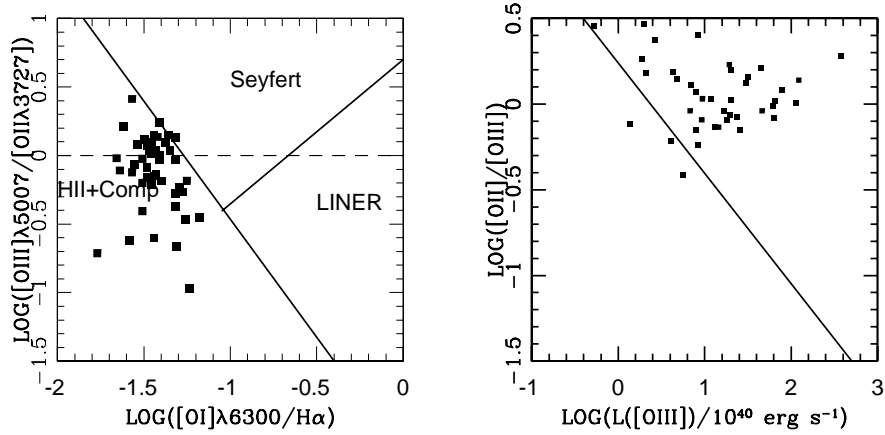}
\caption{\it Left panel: \rm The [\ion{O}{3}]/[\ion{O}{2}] vs. [\ion{O}{1}]/H$\alpha$ 
diagnostic diagram for the composite AGNs. The 
solid lines represent the classification scheme proposed by Kewley et al. (2006), 
and the dashed line the LINER separation line in Heckman (1980). \it Right panel: \rm 
The line ratio [\ion{O}{2}]/[\ion{O}{3}] plotted against the luminosity of [\ion{O}{3}]. 
The solid line shows the least-square regression for type I AGNs given by Kim et al. (2006).
The composite AGNs clearly show enhanced line ratio [\ion{O}{2}]/[\ion{O}{3}].
}  
\end{figure}

\begin{figure}
\includegraphics[width = 13cm]{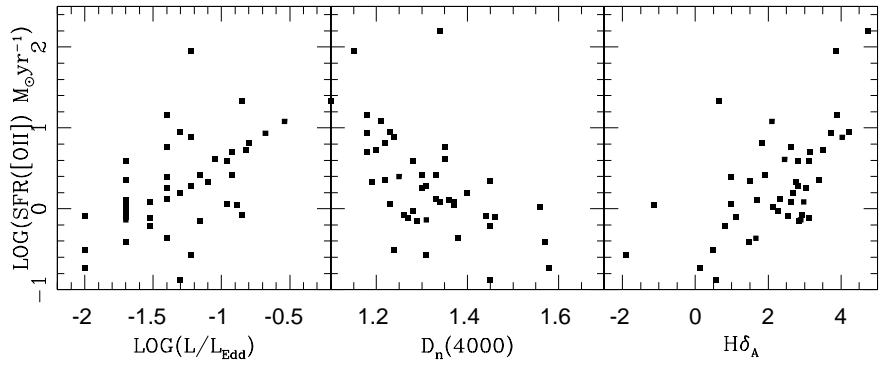}
\caption{[The \ion{O}{2}] luminosity inferred SFR plotted against $L/L_{\mathrm{Edd}}$ (\it 
left panel\rm ), $D_n(4000)$ (\it middle panel\rm), and $H\delta_A$ (\it right panel\rm) for the composite 
AGNs.}
\end{figure}









\begin{deluxetable}{cccccccrcccr}
\tabletypesize{\scriptsize}
\rotate
\tablecaption{List of properties of the composite AGNs from SDSS}
\tablewidth{0pt}
\tablehead{
\colhead{SDSS name} & \colhead{z} & \colhead{[\ion{N}{2}]/H$\alpha$} & \colhead{[\ion{S}{2}]/H$\alpha$} &
\colhead{[\ion{O}{1}]/H$\alpha$} &
\colhead{[\ion{O}{3}]/H$\beta$} & \colhead{$D_n(4000)$} & \colhead{H$\delta_A$} &
\colhead{L(H$\alpha$)} & \colhead{$\log(M/M_{\odot})$} &
\colhead{$L/L_{\mathrm{Edd}}$} & \colhead{L([\ion{O}{3}])}
\\
\colhead{} & \colhead{} & \colhead{} & \colhead{} & \colhead{} & \colhead{} & \colhead{} & \colhead{(\AA)} & 
\colhead{($10^{42}\ \mathrm{erg\ s^{-1}}$)} & \colhead{} & \colhead{} & \colhead{($10^{40}\ \mathrm{erg\ s^{-1}}$)} \\
\colhead{(1)} & \colhead{(2)} & \colhead{(3)} & \colhead{(4)} & \colhead{(5)} & \colhead{(6)} & 
\colhead{(7)} & \colhead{(8)} & \colhead{(9)} & \colhead{(10)} & \colhead{(11)} & \colhead{(12)}\\
}
\startdata
SDSS\,J004311.60-093816.0 & 0.054 & 0.69 & 0.36 & 0.053 & 1.06 & 1.38 & 1.659 & 0.08 & 6.69 & 0.04 & 1.89\\
SDSS\,J004743.93-094641.1 & 0.100 & 0.50 & 0.26 & 0.031 & 0.96 & 1.28 & 2.829 & 0.45 & 6.90 & 0.11 & 19.97\\
SDSS\,J005200.56+003549.3 & 0.114 & 0.49 & 0.27 & 0.044 & 2.00 & 1.29 & 2.836 & 0.18 & 6.73 & 0.07 & 8.02\\
SDSS\,J005312.68-084809.8 & 0.079 & 0.60 & 0.34 & 0.040 & 0.91 & 1.57 & 1.473 & 0.09 & 7.13 & 0.02 & 2.06\\
SDSS\,J014019.05-092110.4 & 0.135 & 0.42 & 0.18 & 0.036 & 0.68 & 1.26 & 2.929 & 0.52 & 6.83 & 0.14 & 4.83\\
SDSS\,J022608.19-005318.7 & 0.106 & 0.32 & 0.29 & 0.046 & 3.62 & 1.14 & 1.364 & 0.33 & 6.66 & 0.14 & 30.95\\
SDSS\,J023140.97-011003.6 & 0.054 & 0.57 & 0.98 & 0.420 & 2.30 & 1.85 & -1.130 & 0.04 & 7.56 & 0.003 & 0.75\\
SDSS\,J033458.00-054853.2 & 0.018 & 0.24 & 0.54 & 0.192 & 1.73 & 1.41 & 1.191 & 0.01 & 6.41 & 0.01 & 0.13\\
SDSS\,J074630.60+440433.5 & 0.193 & 0.41 & 0.29 & 0.037 & 1.76 & 1.10 & 0.649 & 1.89 & 7.30 & 0.14 & 123.35\\
SDSS\,J075352.72+264900.5 & 0.081 & 0.57 & 0.64 & 0.145 & 2.73 & 1.54 & -0.360 & 0.34 & 7.70 & 0.01 & 37.64\\
SDSS\,J080752.27+383211.0 & 0.067 & 0.61 & 0.29 & 0.034 & 1.35 & 1.31 & 2.805 & 0.48 & 7.14 & 0.06 & 16.95\\
SDSS\,J081218.83+362620.5 & 0.122 & 0.52 & 0.35 & 0.045 & 1.64 & 1.27 & 3.136 & 0.27 & 7.30 & 0.03 & 6.81\\
SDSS\,J081509.80+032753.6 & 0.051 & 0.46 & 0.26 & 0.027 & 2.42 & 1.31 & -1.890 & 0.17 & 6.75 & 0.06 & 5.59\\
SDSS\,J084716.44+315309.2 & 0.138 & 0.53 & 0.29 & 0.031 & 0.50 & 1.33 & 5.008 & 0.99 & 7.14 & 0.12 & 8.36\\
SDSS\,J085338.27+033246.1 & 0.208 & 0.56 & 0.20 & \dotfill & 0.22 & 1.16 & 7.224 & 2.59 & 7.04 & 0.34 & 1.48\\
SDSS\,J085824.22+520541.1 & 0.090 & 0.59 & 0.34 & 0.056 & 1.38 & 1.44 & 2.545 & 0.07 & 7.18 & 0.01 & 4.27\\
SDSS\,J090012.02+450514.8 & 0.088 & 0.25 & 0.11 & 0.017 & 0.86 & 1.34 & 4.727 & 1.15 & \dotfill & \dotfill & 244.92\\
SDSS\,J090615.53+463619.0 & 0.085 & 0.70 & 2.47 & 1.629 & 2.00 & 1.99 & -2.77 & 0.25 & 6.70 & 0.10 & 4.36\\
SDSS\,J090837.08+305927.2 & 0.063 & 0.53 & 0.31 & 0.031 & 1.76 & 1.30 & 1.917 & 0.24 & 6.82 & 0.07 & 20.05\\
SDSS\,J093327.87+093200.0 & 0.050 & 0.66 & 0.49 & 0.081 & 1.81 & 1.38 & -1.350 & 0.07 & 6.75 & 0.03 & 0.93\\
SDSS\,J093628.89+475451.4 & 0.055 & 0.51 & 0.42 & \dotfill & 1.86 & 1.62 & 1.093 & 0.11 & 6.94 & 0.03 & 4.02\\
SDSS\,J095030.53+011313.7\tablenotemark{a} & 0.130 & 0.53 & 0.26 & 0.035 & 1.57 & 1.30 & 3.041 & 0.28 & 7.14 & 0.04 & 18.19\\
SDSS\,J095030.53+011313.7\tablenotemark{b} & 0.130 & 0.56 & 0.22 & 0.033 & 1.73 & 1.25 & 0.994 & 0.51 & 7.40 & 0.04 & 23.65\\
SDSS\,J095227.78+014121.3 & 0.074 & 0.44 & 0.29 & 0.035 & 1.32 & 1.23 & 4.223 & 0.43 & 7.22 & 0.05 & 44.58\\
SDSS\,J095558.81+580837.1 & 0.141 & 0.53 & 0.28 & 0.026 & 2.46 & 1.34 & 1.439 & 0.80 & 7.54 & 0.04 & 40.88\\
SDSS\,J100710.35+113146.1 & 0.082 & 0.61 & 0.28 & 0.048 & 1.45 & 1.34 & 2.615 & 0.25 & 7.34 & 0.02 & 13.19\\
SDSS\,J100720.56+562002.9 & 0.130 & 0.53 & 0.36 & 0.037 & 2.00 & 1.20 & 3.494 & 0.63 & 6.88 & 0.15 & 46.30\\
SDSS\,J101628.43+375912.4 & 0.133 & 0.59 & \dotfill & \dotfill & 0.76 & 1.26 & 3.202 & 0.26 & 6.99 & 0.05 & 6.84\\
SDSS\,J102554.66+471424.5 & 0.063 & 0.55 & 0.47 & 0.076 & 2.60 & 1.47 & 1.207 & 0.09 & 6.75 & 0.04 & 5.78\\
SDSS\,J102649.95+402128.9 & 0.063 & 0.55 & 0.40 & 0.049 & 0.57 & 1.56 & 2.142 & 0.07 & 6.94 & 0.02 & 1.80\\
SDSS\,J104643.93+623816.0 & 0.132 & 0.47 & 0.26 & 0.027 & 1.18 & 1.18 & 3.142 & 0.37 & 6.78 & 0.12 & 30.10\\
SDSS\,J105938.92+455329.7 & 0.062 & 0.44 & 0.28 & 0.023 & 1.20 & 1.37 & -1.130 & 0.21 & 6.53 & 0.13 & 6.92\\
SDSS\,J110126.46+094720.0 & 0.027 & 0.54 & 0.73 & 0.091 & 0.71 & 1.38 & -0.180 & 0.09 & \dotfill & \dotfill & 2.15\\
SDSS\,J110654.45+061213.0 & 0.043 & 0.46 & 1.47 & 0.598 & 1.65 & 1.63 & -1.180 & 0.11 & 7.36 & 0.01 & 0.51\\
SDSS\,J110720.51+502935.2 & 0.105 & 0.48 & 0.26 & 0.042 & 1.09 & 1.28 & 2.283 & 0.50 & 7.56 & 0.02 & 9.25\\
SDSS\,J112005.32+545602.5 & 0.072 & 0.36 & 0.77 & 0.865 & 2.24 & 1.86 & -1.970 & 0.10 & 6.81 & 0.03 & 0.95\\
SDSS\,J112455.86+080615.3 & 0.070 & 0.47 & 0.31 & 0.038 & 2.07 & 1.33 & 2.337 & 0.18 & 6.94 & 0.04 & 14.57\\
SDSS\,J112852.59-032130.5 & 0.198 & 0.52 & 0.24 & 0.039 & 1.70 & 1.18 & 3.899 & 2.09 & 7.91 & 0.04 & 114.84\\
SDSS\,J113042.75-004903.0 & 0.105 & 0.58 & 0.36 & 0.048 & 1.62 & 1.36 & 1.686 & 0.21 & 7.25 & 0.02 & 9.44\\
SDSS\,J113258.06+463755.2 & 0.179 & 0.38 & 0.23 & \dotfill & 0.97 & 1.35 & 3.016 & 0.68 & 7.14 & 0.09 & 10.83\\
SDSS\,J114530.25+094344.7 & 0.021 & 0.45 & 0.59 & 0.361 & 0.98 & 2.00 & -2.270 & 0.01 & 6.28 & 0.02 & 0.12\\
SDSS\,J115038.18-005326.6 & 0.051 & 0.54 & 0.46 & 0.055 & 0.99 & 1.31 & 2.890 & 0.05 & 6.87 & 0.02 & 1.98\\
SDSS\,J115226.76+120951.7 & 0.105 & 0.73 & 0.44 & \dotfill & 2.41 & 1.48 & 1.017 & 0.16 & 7.00 & 0.03 & 2.67\\
SDSS\,J120234.39+544500.6 & 0.050 & 0.46 & 0.48 & 0.126 & 2.33 & 1.56 & 0.792 & 0.03 & 6.87 & 0.01 & 0.68\\
SDSS\,J120349.20+020556.9 & 0.081 & 0.57 & 0.32 & 0.039 & 1.73 & 1.45 & 0.805 & 0.23 & 7.14 & 0.03 & 8.39\\
SDSS\,J122926.42+132020.5 & 0.152 & 0.46 & 0.24 & 0.029 & 1.80 & 1.22 & 1.834 & 1.56 & 7.19 & 0.16 & 63.15\\
SDSS\,J123503.48+045530.5 & 0.084 & 0.48 & 0.30 & 0.034 & 1.77 & 1.19 & 2.753 & 0.71 & 7.20 & 0.08 & 19.72\\
SDSS\,J124446.01+602548.7 & 0.151 & 0.40 & 0.25 & 0.022 & 1.20 & 1.18 & 3.711 & 1.77 & 7.10 & 0.21 & 65.66\\
SDSS\,J125400.20+424220.0 & 0.054 & 0.92 & \dotfill & 0.235 & 3.43 & 1.70 & -1.060 & 0.05 & 6.59 & 0.03 & 0.99\\
SDSS\,J130620.96+531823.2 & 0.024 & 1.02 & 1.61 & 0.463 & 1.66 & 1.64 & -0.100 & 0.04 & 7.43 & 0.004 & 0.09\\
SDSS\,J133215.56+585206.1 & 0.142 & 0.44 & 0.32 & 0.082 & 0.76 & 1.21 & 3.474 & 0.41 & 7.60 & 0.02 & 9.64\\
SDSS\,J133715.92+030936.5 & 0.192 & 0.51 & 0.27 & 0.051 & 0.66 & 1.35 & 2.453 & 1.83 & 7.52 & 0.09 & 19.21\\
SDSS\,J134417.23+032633.1 & 0.115 & 1.22 & 0.35 & 0.137 & 6.80 & 1.47 & -0.31 & 0.38 & 7.02 & 0.07 & 17.79\\
SDSS\,J135244.41+572442.3 & 0.056 & 0.23 & 0.41 & 0.076 & 2.74 & 1.40 & 1.954 & 0.06 & 6.74 & 0.03 & 2.69\\
SDSS\,J135719.47+394045.3 & 0.265 & 0.44 & 0.24 & \dotfill & 0.93 & 1.18 & 4.319 & 5.06 & 7.70 & 0.13 & 86.13\\
SDSS\,J135910.91+531933.7 & 0.075 & 0.41 & 0.28 & 0.024 & 1.61 & 1.24 & 0.497 & 0.23 & 7.60 & 0.01 & 4.11\\
SDSS\,J140941.69+372905.0 & 0.036 & 0.46 & 0.61 & 0.066 & 1.13 & 1.58 & 0.133 & 0.03 & 6.81 & 0.01 & 0.52\\
SDSS\,J142644.43+640703.7 & 0.132 & 0.43 & 0.31 & 0.056 & 3.06 & 1.17 & 3.911 & 0.66 & 6.54 & 0.33 & 40.84\\
SDSS\,J145054.16+350837.8 & 0.206 & 0.60 & 0.27 & 0.048 & 1.33 & 1.15 & 3.847 & 4.63 & 7.99 & 0.06 & 374.13\\
SDSS\,J145310.46+580955.5 & 0.112 & 0.51 & 0.38 & \dotfill & 0.75 & 1.31 & 4.229 & 0.97 & 7.15 & 0.11 & 17.10\\
SDSS\,J150307.19+421744.1 & 0.126 & 0.79 & \dotfill & \dotfill & 1.93 & 1.41 & -0.04 & 0.40 & 6.93 & 0.09 & 3.91\\
SDSS\,J150803.17+485728.1 & 0.111 & 0.41 & 0.23 & 0.033 & 1.11 & 1.35 & 2.638 & 1.59 & 7.74 & 0.04 & 32.03\\
SDSS\,J150829.43+371120.8 & 0.055 & 0.76 & 0.52 & 0.058 & 0.67 & 1.45 & 1.504 & 0.04 & \dotfill & \dotfill & 1.88\\
SDSS\,J151014.20-000135.0 & 0.093 & 0.57 & 0.48 & \dotfill & 2.20 & 1.41 & 1.477 & 0.18 & 7.06 & 0.03 & 9.51\\
SDSS\,J151732.81+602844.6 & 0.129 & 0.86 & 0.37 & 0.075 & 7.71 & 1.47 & -0.240 & 0.68 & 7.17 & 0.08 & 45.87\\
SDSS\,J152023.49+502947.8 & 0.093 & 0.42 & 0.31 & 0.026 & 0.59 & 1.28 & 3.113 & 0.43 & 7.58 & 0.02 & 7.48\\
SDSS\,J152036.05+312225.8 & 0.107 & 0.51 & 0.31 & 0.047 & 2.29 & 1.44 & 0.943 & 0.94 & 8.05 & 0.01 & 21.32\\
SDSS\,J153015.55+531035.1 & 0.174 & 0.43 & 0.26 & 0.035 & 1.85 & 1.24 & 4.034 & 1.39 & 7.56 & 0.06 & 62.76\\
SDSS\,J163330.21+213634.9 & 0.095 & 0.52 & 0.57 & 0.176 & 2.28 & 1.84 & -2.150 & 0.18 & 7.28 & 0.02 & 12.14\\
SDSS\,J164643.66+201654.5 & 0.081 & 0.64 & 0.36 & 0.048 & 1.11 & 1.46 & 1.113 & 0.27 & 7.34 & 0.02 & 2.65\\
SDSS\,J163413.74+445947.8 & 0.058 & 0.55 & 0.38 & 0.058 & 2.46 & 1.38 & 1.427 & 0.17 & 6.83 & 0.05 & 8.65\\
SDSS\,J164531.50+403224.6 & 0.112 & 0.50 & 0.31 & 0.051 & 1.99 & 1.26 & 3.600 & 0.32 & 7.16 & 0.04 & 32.95\\
SDSS\,J164552.20+442614.8 & 0.070 & 1.37 & 1.06 & 0.511 & 5.74 & 1.50 & -1.040 & 0.61 & \dotfill & \dotfill & 5.04\\
SDSS\,J165039.15+634429.0 & 0.148 & 0.42 & 0.27 & 0.036 & 1.42 & 1.22 & 3.390 & 0.50 & 7.57 & 0.02 & 25.31\\
SDSS\,J171855.43+275947.4 & 0.111 & 0.49 & 0.28 & 0.039 & 1.34 & 1.40 & 2.684 & 0.36 & 7.13 & 0.05 & 11.94\\
SDSS\,J210641.11-075501.2 & 0.131 & 0.64 & 0.48 & 0.080 & 1.13 & 1.38 & 2.289 & 0.15 & 6.29 & 0.17 & 9.77\\
SDSS\,J210721.91+110359.0 & 0.042 & 0.26 & 0.47 & 0.103 & 2.52 & 1.47 & 0.996 & 0.06 & 6.55 & 0.04 & 1.31\\
SDSS\,J213741.44+005316.9 & 0.041 & 1.16 & \dotfill & 0.081 & 3.23 & 1.58 & -1.180 & 0.04 & 6.50 & 0.03 & 0.72\\
SDSS\,J220531.56-073734.8 & 0.060 & 0.85 & 0.77 & 0.161 & 3.18 & 1.41 & -0.500 & 0.09 & 7.35 & 0.01 & 0.73\\
SDSS\,J222705.69-090424.8 & 0.073 & 0.49 & 0.30 & 0.032 & 1.25 & 1.45 & 0.568 & 0.10 & 6.63 & 0.05 & 1.36\\
SDSS\,J225057.27-085410.9 & 0.065 & 0.45 & 0.41 & 0.045 & 2.54 & 1.28 & 2.756 & 0.08 & 7.00 & 0.02 & 17.22\\
SDSS\,J225102.65-093824.1 & 0.081 & 0.40 & 0.40 & 0.036 & 0.65 & 1.37 & 2.976 & 0.09 & 6.81 & 0.03 & 2.42\\
SDSS\,J232643.08-095927.8 & 0.070 & 0.42 & 0.36 & 0.052 & 4.03 & 1.25 & -0.22 & 0.13 & 6.51 & 0.09 & 6.00\\
SDSS\,J232714.52-102317.5 & 0.065 & 0.52 & 0.30 & 0.028 & 0.95 & 1.23 & 0.991 & 0.18 & 6.55 & 0.11 & 7.92\\
SDSS\,J233942.54+142305.3 & 0.149 & 0.49 & 0.22 & 0.033 & 1.69 & 1.21 & 2.105 & 1.69 & 6.95 & 0.29 & 79.06\\

\enddata
\tablenotetext{a} {The spectra ID is \sl spSpec51608-0267-405 \rm given by SDSS.}
\tablenotetext{b} {The spectra ID is \sl spSpec51989-0480-120 \rm given by SDSS.}
\end{deluxetable}


\clearpage



\clearpage




\end{document}